\documentclass[conference,hyphens]{IEEEtran}
\usepackage{amsmath,amsfonts}
\usepackage{algorithmic}
\usepackage{graphicx}
\usepackage{textcomp}
\usepackage{xcolor}
\usepackage{booktabs}
\usepackage{tagging}
\usepackage{tikz}
\usepackage{amsmath}
\usepackage{float}
\usepackage[binary-units]{siunitx}
\DeclareSIUnit\event{event}
\usepackage{times,url,color,soul,xspace}
\usepackage[inline]{enumitem}
\usepackage{graphicx}
\usepackage{caption}
\usepackage{subcaption}
\usepackage{comment}
\usepackage{xspace}
\usepackage{hyperref}
\usepackage[inline,draft,nomargin,index]{fixme}
\usepackage{cprotect}
\usepackage{wrapfig,booktabs}
\usepackage[binary-units]{siunitx}
\setlist{nolistsep,leftmargin=*}
\usepackage{orcidlink} 
\usepackage{cleveref}
\usepackage{multirow}
\crefformat{section}{\S#2#1#3}
\crefformat{subsection}{\S#2#1#3}
\crefformat{subsubsection}{\S#2#1#3}
\definecolor{codegreen}{rgb}{0,0.6,0}
\definecolor{codegray}{rgb}{0.5,0.5,0.5}
\definecolor{codepurple}{rgb}{0.58,0,0.82}
\definecolor{backcolour}{rgb}{0.95,0.95,0.92}
\usepackage{listings}
\lstdefinestyle{sqlstyle}{
    language=SQL,
    basicstyle=\ttfamily\small,
    keywordstyle=\color{blue}\bfseries,
    stringstyle=\color{teal},
    commentstyle=\color{gray},
    frame=single,
    framerule=0.1pt,
    rulecolor=\color{black},
    backgroundcolor=\color{backcolour},
    showstringspaces=false,
    breaklines=true,
}
\lstdefinestyle{terminal}{
  basicstyle=\ttfamily\footnotesize,
  backgroundcolor=\color{backcolour},
  commentstyle=\color{codegreen},
  keywordstyle=\color{magenta},
  numberstyle=\tiny\color{codegray},
  stringstyle=\color{codepurple},
  breaklines=true,
  postbreak=\mbox{\textcolor{red}{$\hookrightarrow$}\space},
  showstringspaces=false,
  keywordstyle=\color{blue},
  commentstyle=\color{green},
  stringstyle=\color{red},
  basicstyle=\ttfamily\color{black}
}

\FXRegisterAuthor{giu}{agiu}{\textcolor{red}{Giu}}

\FXRegisterAuthor{atr}{aatr}{\textcolor{blue}{Atr}}

\FXRegisterAuthor{lin}{alin}{\textcolor{green}{Lin}}

\newcommand{\upp}{\vspace*{-0.5em}}

\makeatletter

\newcommand*\circled[1]{%
  \tikz[baseline=(C.base)]\node[draw,circle,inner sep=0.6pt,line width=0.2mm,](C) {#1};}

\usepackage[table]{xcolor}
\usepackage{colortbl}

\definecolor{bestgreen}{RGB}{0,150,0}
\definecolor{goodgreen}{RGB}{120,200,120}
\definecolor{mediocre}{RGB}{255,220,150}
\definecolor{badred}{RGB}{255,150,150}

\usepackage{fancyhdr}
\def\BibTeX{{\rm B\kern-.05em{\sc i\kern-.025em b}\kern-.08em
    T\kern-.1667em\lower.7ex\hbox{E}\kern-.125emX}}

\begin{document}
\title{Seeing Through NAT with Ripple: Retrofitting Service Dependency Discovery in Networked Distributed Systems}


% \author{\IEEEauthorblockN{Diogo Landau\orcidlink{0009-0000-2625-8884}\IEEEauthorrefmark{1},
% \author{\IEEEauthorblockN{Gijs Blanken\orcidlink{0009-0005-5633-1778}\IEEEauthorrefmark{1},
% Jorge Barbosa\orcidlink{0000-0003-4135-2347}\IEEEauthorrefmark{2}, 
% Nishant Saurabh\orcidlink{0000-0002-1926-4693}\IEEEauthorrefmark{1}}}}
% \IEEEauthorblockA{\IEEEauthorrefmark{1}\textit{Department of Information and Computing Sciences, Utrecht University, NL}}
% \IEEEauthorrefmark{2}{\textit{LIACC, Faculdade de Engenharia da Universidade do Porto, Portugal}}\\
% d.hewittmouracarrapatolandau@uu.nl\IEEEauthorrefmark{1},blankengijs@gmail.com\IEEEauthorrefmark{1},
% jbarbosa@fe.up.pt\IEEEauthorrefmark{2}
% n.saurabh@uu.nl\IEEEauthorrefmark{1}
% \upp\upp\upp
% }

\author{\IEEEauthorblockN{Diogo Landau\orcidlink{0009-0000-2625-8884
}\IEEEauthorrefmark{1}, Gijs Blanken\orcidlink{0000-0002-1926-4693}\IEEEauthorrefmark{1},
Jorge Barbosa\orcidlink{0000-0003-4135-2347}\IEEEauthorrefmark{2}, Nishant Saurabh\orcidlink{0000-0002-1926-4693}\IEEEauthorrefmark{1}} 
\IEEEauthorblockA{\IEEEauthorrefmark{1}\textit{Department of Information and Computing Sciences, Utrecht University, NL}}
\IEEEauthorrefmark{2}{\textit{LIACC, Faculdade de Engenharia da Universidade do Porto, Portugal}}\\
d.hewittmouracarrapatolandau@uu.nl\IEEEauthorrefmark{1}, blankengijs@gmail.com\IEEEauthorrefmark{1},
jbarbosa@fe.up.pt\IEEEauthorrefmark{2},
n.saurabh@uu.nl\IEEEauthorrefmark{1}
\upp\upp\upp
}

\maketitle

\begin{abstract}
  Modern distributed systems rely on complex networks of interconnected services, creating direct or indirect dependencies that can propagate faults and cause cascading failures. To localize the root cause of performance degradation in these environments, constructing a \textit{service dependency graph} is highly beneficial. However, building an accurate service dependency graph is impaired by complex routing techniques, such as Network Address Translation (NAT), an essential mechanism for connecting services across networks. NAT obfuscates the actual hosts running the services, causing existing run-time approaches that passively observe network metadata to fail in accurately inferring service dependencies. 
  
  To this end, this paper introduces Ripple, a run-time system for constructing process-level service dependency graphs. It operates without source code instrumentation and remains resilient under complex network routing mechanisms, including NAT. Ripple implements a non-disruptive method of injecting metadata onto a TCP packet's header that maintains protocol correctness across host boundaries. In other words, if no receiving agent is present, the instrumentation leaves existing TCP connections unaffected, ensuring non-disruptive operation when it is partially deployed across hosts. We evaluated Ripple extensively against three state-of-the-art systems across nine scenarios, involving three network configurations (NAT-free, internal-NAT, external-NAT) and three microservice benchmarks. Ripple was the only approach that performed consistently across networking configurations. With regards to correctness, it performed on par with, or better than, the state-of-the-art with precision and recall values of 100\% in the majority of the scenarios.
\end{abstract}



\maketitle

\section{Introduction}
Large-scale networked distributed systems are a de-facto standard when developing applications with stringent non-funtional requirement guarantees, such as, availability, scalability and performance~\cite{koponen2010onix,wang2015exploring}. In such environments, services are a basic building block providing data and functionality to communicating processes. However, when a critical service's performance degrades, it may lead to cascading failures~\cite{ashok2024traceweaver,zhang2022crisp} that impair end-user experience, often in the form of increased latency.
% Large-scale networked distributed systems are a de-facto standard when developing applications with stringent non-funtional requirement guarantees, such as, availability, scalability and performance~\cite{koponen2010onix,wang2015exploring}. Services coordinate within such environments to ensure critical functionality, but their coordination can trigger cascading failures if one service’s performance degrades~\cite{ashok2024traceweaver,zhang2022crisp}.
% The coordination between services in such an environment is critical to satisfy functionality, however, it may also lead to cascading failure when a service's performance degrades \cite{ashok2024traceweaver,zhang2022crisp}. 
In other words, response time increase observed in a particular service may be the result of degradation experienced in another service it directly or indirectly depends on. Consequently, given the complex interdependencies common in distributed systems, localizing the root cause of performance degradation can benefit from utilizing a \textit{service dependency graph} \cite{harsh2023murphy,gan2021sage,he2022graph,huye2023lifting}, which narrows the search space to only those services involved in communication during the observed degradation timeframe.

Unfortunately, the complex network routing methods used in modern deployments hinder the construction of accurate service dependency graphs. Network Address Translation (NAT), while invaluable for connecting services across different networks, obscures the actual host running the service involved in the communication. Containerization illustrates a typical configuration where NAT operates at the container’s host \cite{suo2018analysis}, while cross Local Area Network (LAN) communication relies on NAT at the network gateway node \cite{wing2010network}. 
% Common network configurations that rely on NAT include: containerization where NAT occurs at the container’s host~\cite{suo2018analysis}; and cross-LAN communication, where NAT is performed at the network’s gateway node \cite{wing2010network}. 
Similarly, Virtual Private Networks (VPNs) enable connecting hosts using private IP address ranges~\cite{zhuo2019slim} across different LANs. 
Packet encapsulation, which enables such connectivity, introduces additional complexity, complicating the resolution of the services participating in communication and further impairing accurate service dependency inference.
% Similarly, Virtual Private Networks (VPNs) may also be used to connect hosts using private IP address ranges~\cite{zhuo2019slim}, even when these hosts are located in different Local Area Networks. The packet encapsulation that enables this connectivity adds a layer of complexity that complicates the resolution of the services participating in the communication, thereby impairing service dependency inference.

Several approaches exist to uncover service dependencies in distributed systems, each operating at a different stage of an application's lifecycle. Build-time methods \cite{saokar2023servicerouter,sigelman2010dapper,zipkin,jaeger,opentelemetry} provide high-fidelity observability by embedding tracing logic directly into the application’s source code. However, these methods require substantial developer effort and ongoing maintenance, and are unsuitable for third-party or legacy services. Deploy-time techniques \cite{envoy,istio,linkerd}, popularised by service meshes, mitigate these challenges by intercepting application traffic through sidecar proxies. Although this model simplifies adoption compared to build-time instrumentation, recent studies \cite{zhu2023dissecting} demonstrate it introduces significant CPU and latency overheads. The model also mandates that engineers deploy all services with a compatible sidecar, which makes retrofitting it into existing architectures difficult.

To enable retrofitting observability onto existing deployments, run-time approaches~\cite{shen2023network,caretta,coroot} infer dependencies by passively observing network metadata exchanged between services (e.g., source, destination IP addresses). However, the presence of NAT impedes these methods. Another run-time approach \cite{yang2025zerotracer} injects trace identifiers into HTTP headers when services communicate via plaintext HTTP connections. However, its inability to operate over encrypted channels (e.g., HTTPS) or to support protocols beyond HTTP underscores the need for a more general solution, which can remain effective in encrypted environments and accommodate a broader range of stateful communication protocols.

To this end, this paper presents Ripple, a method for constructing a distributed system’s service dependency graph that is resilient to the complex routing mechanisms employed in modern deployments. Ripple works by injecting an identifier using eBPF~\cite{eBPF} into a single packet of a Transmission Control Protocol (TCP) connection on a host’s egress path, which can then be intercepted at receiving host’s ingress path. Crucially, if no agent is present on receiving host to capture the identifier, the design of our approach ensures that connection between services remains unaffected. In addition to identifying the communicating network hosts, Ripple goes a step further by resolving specific processes involved in the communication on each host. This makes it suitable for service deployments in both containerized and native host-level environments.

To evaluate Ripple, we applied it to three standard microservice benchmark applications~\cite{gan2019open,online-boutique}, each comprising between $16$ and $34$ service dependencies, and tested it across three distinct network configurations (i.e., NAT-free, internal-NAT, external-NAT) to assess its robustness to varying routing methods with respect to accuracy and recall. Our results show that Ripple performs on par with, or better than, state-of-the-art approaches~\cite{caretta,coroot,shen2023network}, while incurring overhead comparable to the least intrusive methods. We also report Ripple’s time-to-completion as a proxy for evaluating its scalability. Our findings indicate that
%, at the scale tested, 
the time required to construct the full service dependency graph is influenced more by the frequency of inter-service communication than by the absolute number of services or dependencies. Finally, we present a motivating use case demonstrating how Ripple’s output can be leveraged to diagnose the root cause of performance degradation in a microservice application composed of $12$ services and $16$ dependencies. Ripple’s complete source code and reproducibility artifact is released opensource in an \emph{online} \verb|GitHub| repository~\cite{repo,repro}. 

\section{Related Work}
\label{sec:related-works}
Understanding service dependencies is fundamental to managing and troubleshooting modern distributed systems, particularly in cloud environments dominated by microservice architectures. Traditional manual approaches to dependency mapping, such as those used by \cite{suo2018vnettracer}, rely on a pre-existing understanding of the system’s architecture. However, as deployments scale and diverge from their initial design, these methods quickly become impractical. While sufficient for small or static systems, manual mapping does not extend to the size and dynamism of real-world environments, where distributed applications may involve thousands of interacting microservices \cite{winchester2025complexity,zhang2022crisp} (ranging from $4,000$ to $300,000$). At such scales, automated observability becomes essential. Next, we categorize state-of-the-art techniques for uncovering service dependencies by instrumentation strategy: (1) \textit{build-time}, (2) \textit{deploy-time}, and (3) \textit{run-time}.

\subsection{Build-time Instrumentation}
Build-time instrumentation involves modifying an application’s source code before it is compiled and packaged for deployment. It offers high accuracy in tracing request flows and is widely used to enhance observability in distributed systems \cite{yang2025zerotracer}. However, it imposes a significant maintenance burden, as developers must explicitly annotate code paths based on their understanding of the system’s architecture and execution flow.

A common build-time technique is distributed tracing \cite{sigelman2010dapper,zipkin,jaeger,opentelemetry,li2025tracewizard}, which records the lifecycle of requests as they traverse multiple services. These systems offer language-specific SDKs that enable developers to integrate tracing into their applications. More recently, advances in automatic instrumentation \cite{saokar2023servicerouter,opentelemetry-zero-code} have introduced support for tracing at the library level, where observability is added implicitly as long as the application uses instrumented libraries. When correctly configured, build-time instrumentation can offer comprehensive and precise observability. 

{\color{black}
Build-time approaches~\cite{sigelman2010dapper,zipkin,jaeger,opentelemetry,li2025tracewizard}, however, require that the services are compiled or configured with tracing libraries to operate, creating a dependency on instrumentation at execution time. Consequently, services lacking such instrumentation (e.g., legacy systems or third-party software) generate no traces. This results in blind spots within the distributed system, hindering performance degradation analysis ~\cite{ashok2024traceweaver}. Furthermore, even a complete deployment of these distributed tracing systems introduces substantial runtime overhead, with throughput degradations ranging from 38\% to 44\%~\cite{nou2025investigating}. Next, we discuss deploy-time and run-time instrumentation strategies which mitigate limitations  of build-time approaches by enabling  observability for heterogeneous, third-party, or legacy services without requiring explicit developer intervention, particularly where source-level instrumentation is infeasible.

% However, build-time approaches~\cite{sigelman2010dapper,zipkin,jaeger,opentelemetry,li2025tracewizard} require services to be compiled or configured with tracing libraries at execution time. Services that are not instrumented accordingly (e.g. legacy code and third-party software), produce no traces, effectively creating blind spots in the distributed system, hindering performance degradation analysis~\cite{ashok2024traceweaver}. Moreover, even when fully deployed, distributed tracing systems introduce substantial runtime overhead, with reported throughput degradations ranging from 38\% to 44\%~\cite{nou2025investigating}. Recognizing the limitations of build-time instrumentation, the following sections discuss deploy- and run-time strategies that aim to provide observability without requiring explicit developer intervention. This makes them more suitable for heterogeneous, third-party, or legacy services where source-level instrumentation is infeasible.
}

\subsection{Deploy-time Instrumentation}
Deploy-time instrumentation introduces observability at deployment phase, typically without requiring changes to the application’s source code. This is achieved by running services alongside sidecar proxies \cite{envoy,istio,linkerd} which intercept and route all traffic between the services in the distributed system, enabling features such as distributed tracing, traffic shaping, and load balancing. This model has gained traction through service mesh architectures, which provide a programmable layer for managing service-to-service communication \cite{ashok2021leveraging,antichi2020full,cao2021co,dab2020cloud,johng2019harmonia}. Nevertheless, these approaches incur significant performance costs, adding up to 92\% extra CPU usage and increasing the latency by up to 185\%~\cite{zhu2023dissecting}. 
Moreover, deploy-time instrumentation require services deployment with the necessary sidecar components, which makes incremental adoption challenging, as any uninstrumented service creates a blind spot in the dependency map.

\subsection{Run-time Instrumentation}
\label{secR:run-time}

Run-time instrumentation techniques infer service dependencies without requiring modifications at build- or deploy-time. This makes them well-suited for retrofitting observability into existing systems, including legacy applications and third-party services where source code is unavailable. This flexibility establishes run-time approaches as the core focus of our work, and we evaluate the proposed Ripple system against tools within this category. Despite these advantages, run-time methods often introduce additional complexity and engineering trade-offs, particularly between breadth of applicability and risk of disruption. To minimise disruption, \textit{observer} approaches passively inspect IP packet contents without interfering with communication. However, as will be presented in Section~\ref{sec:problem-formulation}, this limits their applicability due to in-transit packet metadata modifications. On the other hand, approaches that aim to maximise applicability often opt for \textit{modifying} packet contents to inject an identifier, such that a packet sent by a process can be recognised at the receiving end. While effective, tampering with a packet's contents requires extra care to guarantee it does not compromise an established connection. We categorise the following run-time approaches based on whether they \textit{modify} or simply \textit{observe} a packet's contents.

\textbf{Observers} infer service dependencies by monitoring communication without modifying the data in transit. These approaches analyse traffic patterns, such as network metadata or protocol behavior, to reconstruct the dependencies of a distributed system.

\sloppy
Caretta~\cite{caretta} infers application flows using the packet's five-tuple: $\langle$\texttt{protocol}, \texttt{source IP}, \texttt{source port}, \texttt{destination IP}, \texttt{destination port} $\rangle$. However, Caretta fails to consider Network Address Translation (NAT), a mechanism commonly leveraged in containerised environments \cite{suo2018analysis}, as detailed in Section~\ref{sec:containerisation}. In such settings, NAT mechanisms \cite{wing2010network} frequently change or obfuscate IP addresses and ports, further impairing service dependency discovery accuracy. Coroot \cite{coroot} partially addresses this limitation by inspecting the host’s \texttt{conntrack} table to resolve the original destination after NAT. It enables correct inference of service connections for locally administered NAT. However, external NAT events beyond Coroot’s administrative control cause the strategy to fail, preventing the tool from identifying the actual communicating services.

\sloppy
{\color{black}
Recent works such as Deepflow~\cite{shen2023network}, DeepTrace~\cite{geng2025low}, Traceweaver~\cite{ashok2024traceweaver} leverage temporal correlations and packet-level data context to infer service dependencies in distributed systems. Although effective in small-scale deployments, these approaches suffer from degrading accuracy as service count and parallelism increases, which leads to higher ambiguity in request correlation~\cite{geng2025low}. Moreover, DeepFlow relies on simplified service architecture assumptions that do not generalize effectively to complex, multi-threaded systems, such as those using asynchronous execution models, resulting in incomplete or inaccurate dependency inference. DeepTrace partially mitigates these limitations through deeper packet inspection, enabling richer contextual analysis of service interactions. However, this reliance on payload visibility prevents its use on encrypted traffic~\cite{geng2025low} (e.g., HTTPS), which is ubiquitous in modern distributed systems, thereby limiting its practical applicability.

% Recent works such as \cite{shen2023network,geng2025low,ashok2024traceweaver} exploit temporal correlations and packet-level data context to infer service dependencies in distributed systems. While these approaches are effective in small-scale deployments, their accuracy degrades as the number of services and parallelism increases, leading to higher ambiguity in request correlation \cite{geng2025low}. Moreover, DeepFlow relies on simplified service architecture assumptions that do not generalize well to complex, multi-threaded systems, such as those employing asynchronous execution models, resulting in incomplete or inaccurate dependency inference. DeepTrace partially mitigates these limitations through deeper packet inspection, enabling richer contextual analysis of service interactions. However, this reliance on payload visibility prevents its use on encrypted traffic~\cite{geng2025low} (e.g., HTTPS), which is ubiquitous in modern distributed systems, thereby limiting its practical applicability.
}

Despite avoiding traffic modification complexity, observers often rely on strong assumptions about system architecture and network behavior that may not hold in real-world distributed environments. Hence, their effectiveness is limited in modern systems, where applications adopt diverse architectures and communication patterns.

\textbf{Modifiers} alter a packet's contents in transit to explicitly tag and trace interactions between services, enhancing the accuracy of dependency mapping. ZeroTracer \cite{yang2025zerotracer} exemplifies this approach by propagating trace information via HTTP headers, embedding unique identifiers into requests to link the sending and receiving services. Because these identifiers remain unchanged as packets traverse the network, they provide a reliable means of identifying service relationships. However, ZeroTracer is limited to plaintext HTTP traffic; it cannot inject identifiers into encrypted connections (e.g., HTTPS) or non-HTTP protocols. 
{\color{black}
TraceWizard~\cite{li2025tracewizard}, in contrast, injects trace identifiers into the IPv4 options field, remaining compatible with encrypted traffic. However, previous studies~\cite{gont2014rfc,fransson2004end} have shown that IPv4 packets carrying options frequently suffer from increased delay, jitter and packet loss. Notably, ~\cite{fransson2004end} observed loss rates as high as 99.9\% in 14.3\% of tested environments, suggesting limited practicality in production settings. 

In contrast, leveraging TCP options to propagate additional context is a feasible design choice, as TCP options are explicitly intended to support protocol extensions. Hilal et al.~\cite{yazdani2025poster,hilal2023yarrpbox} measure the survivability of the experimental TCP Options across $103,000$ different paths and show that it was stripped in only $29$ of those paths. In comparison, the standardized SACK TCP Option was stripped in $14$ of those paths.
% Prior measurements across diverse networks including \emph{home}, \emph{hotspot}, \emph{cellular}, \emph{university}, \emph{enterprise}, and \emph{hosting} environments, spanning 142 paths in 24 countries, show minimal interference when packets carry unknown TCP options ~\cite{honda2011still}. Most tested paths forwarded both known and unknown TCP options unmodified, indicating  their suitability for transporting additional context alongside TCP traffic. 
Our tool Ripple builds on these observations, thereby providing protocol-agnostic support over TCP while remaining compatible with encrypted connections. Moreover, unlike TraceWizard~\cite{li2025tracewizard}, which requires services to be compiled or configured with tracing libraries at build-time, Ripple operates through run-time instrumentation and therefore falls within the class of run-time modifier-based approaches.

% However, previous studies~\cite{gont2014rfc,fransson2004end} have shown that IPv4 packets carrying options often experience an increase in delay, jitter and loss rate, with~\cite{fransson2004end} reporting loss rates of 99.9\% in 14.3\% of the tested environments, which may limit the practicality of this approach in some production settings.
% In contrast, leveraging TCP options to propagate additional context is a feasible design choice, as TCP options are explicitly intended to support protocol extensions. Prior measurements~\cite{honda2011still} show minimal interference when packets carry unknown TCP options across a diverse set of network environments, including \emph{home}, \emph{hotspot}, \emph{cellular}, \emph{university}, \emph{enterprise}, and \emph{hosting} networks, spanning 142 network paths across 24 countries. The authors report that the majority of tested paths forward both known and unknown TCP options without modification, indicating their suitability for transporting additional context alongside TCP traffic. Ripple builds on these observations by instrumenting a layer above TraceWizard~\cite{li2025tracewizard} within the TCP options space, thereby providing protocol-agnostic support over TCP while remaining compatible with encrypted connections. Moreover, unlike TraceWizard~\cite{li2025tracewizard}, which requires services to be compiled or configured with tracing libraries at execution time, Ripple operates through run-time instrumentation and therefore falls within the class of run-time modifier-based approaches.
}

\section{Problem Formulation}
\label{sec:problem-formulation}
Given a set of interconnected services deployed in a distributed environment, our goal is to infer a service dependency graph that represents communication between these services. A key challenge in this process is Network Address Translation (NAT), which obscures the actual hosts involved in the communication. NAT modifies packet headers, such as the destination IP address, to enable routing based on a different set of rules than those applied before translation \cite{tsuchiya1991ip,olteanu2015lost}. A NAT rule applies to packets that match specific criteria. 
% For example, Listing~\ref{lst:docker-proxy} shows an iptables Destination NAT (DNAT) rule that rewrites packets with destination port 5000 to have the new destination 172.16.0.2:1234. 
In many setups, Destination NAT (DNAT) accompanies Source NAT (SNAT) to ensure that return traffic routes correctly back through the NAT device, thus maintaining connection consistency for both communicating parties. Given the complexity of modern network configurations involving NAT, this section outlines typical networking scenarios and discusses how they complicate efforts to reconstruct accurate service dependency graphs using existing run-time instrumentation approaches (see Section~\ref{secR:run-time}).

\subsection{Networking in NAT-Free Environments}
\label{sec:nonat}
\begin{figure}[t]
\centering
\includegraphics[width=\columnwidth]{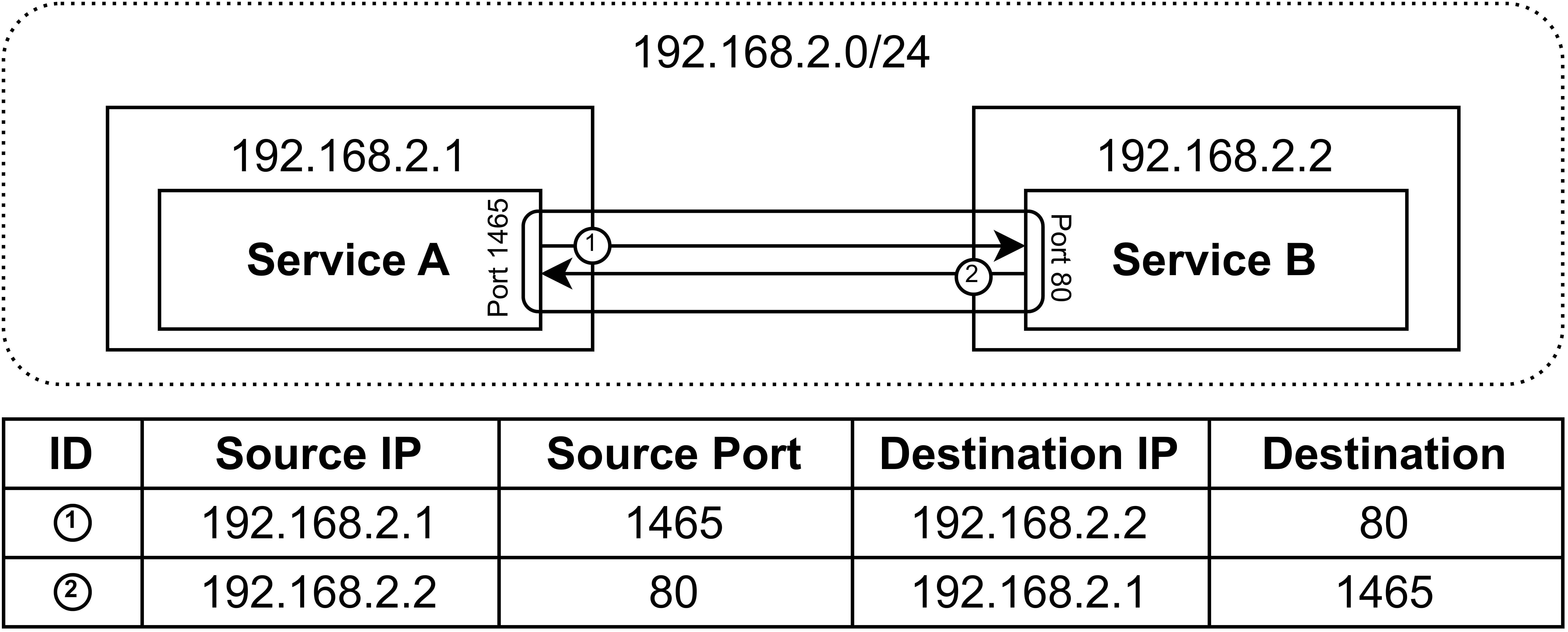}
%\upp
\caption{\textbf{NAT-free network configuration.} Services A and B run on separate hosts within the same network and communicate over a direct connection.}
\label{fig:nat-free} 
%\upp
\end{figure}

When a packet is routed between hosts without any NAT, inspecting its IP addresses and TCP ports at any point along its path is sufficient to identify the communicating hosts. Figure~\ref{fig:nat-free} illustrates this configuration, where services A and B run on hosts within the same network and communicate directly using their respective IP addresses. A packet leaving Service A's host (i.e., \texttt{192.168.2.1}) contains the necessary routing data for correct delivery to Service B running on host \texttt{192.168.2.2}.

%0.8, 0.9

\begin{figure}[t]
\centering
\upp
\includegraphics[width=\columnwidth]{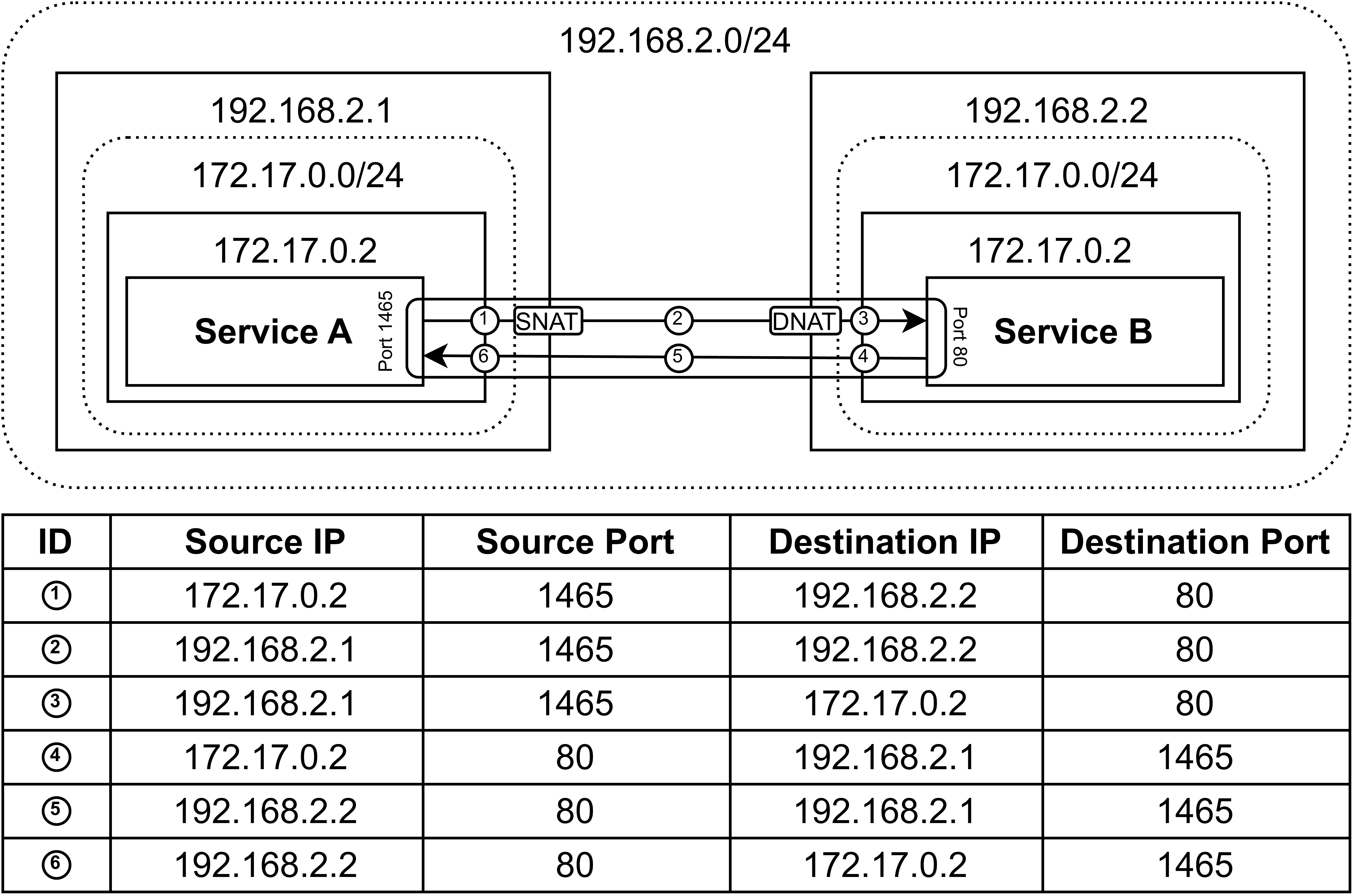}
%\upp
\caption{\textbf{Internal NAT network configuration.} Services A and B run in containers on separate hosts, with NAT applied at each host boundary.}
\label{fig:containerisation} 
\upp\upp
\upp
\end{figure}

An interesting extension of the above example involves Virtual Private Networks (VPNs), where hosts use the VPN's private IP address range to reach other participating hosts in the same network, even when the hosts belong to different Local Area Networks (LANs). Examples include container overlay networks \cite{nam2020bastion,flannel,calico,dockeroverlay} and encrypted tunnels \cite{donenfeld2017wireguard,openvpn,ramesh2023all}. The VPN encapsulates packets to enable routing to the intended physical host before decapsulation and normal routing proceeds. Under these conditions, state-of-the-art approaches for constructing service dependency graphs \cite{caretta,coroot,shen2023network} perform effectively and accurately identify the hosts running the communicating services.

\subsection{Containerisation's Impact on Networking}
\label{sec:containerisation}
Containerisation has become the standard deployment model in modern systems due to its strong isolation and environment consistency guarantees \cite{zhuo2019slim}. On Linux, one of the key features enabling container isolation is the use of network namespaces \cite{jarkas2025container}, which provide each container with its own independent networking stack, separate from that of the host.

To enable communication between containers and the outside world, the most common setup creates a bridge interface in the host's network namespace \cite{suo2018analysis}. Each container is connected to the bridge via a virtual Ethernet (veth) pair, placing one end in the container's namespace and the other in the host's. Containers attached to the same bridge 
%consequently 
share a private IP address range that is reachable from the host. Internal communication between containers on the same host, using this private IP range, does not require NAT. However, when a container attempts to initiate a connection to a service on a different host, SNAT becomes necessary to ensure that responses can be correctly routed back to the originating container.

Figure~\ref{fig:containerisation} illustrates a scenario where Service A (container IP \texttt{172.17.0.2}, host \texttt{192.168.2.1}) initiates a connection to Service B (container IP \texttt{172.17.0.2}, host \texttt{192.168.2.2}) with port 80 exposed (e.g., via Docker’s \texttt{-p 80:80} flag \cite{dockerpublish}). Since containers remain routable only within their hosts, each host assigns identical IP ranges to its container bridge network, where both containers can have identical IP addresses without conflict, while enabling correct cross-host packet routing.
% Figure~\ref{fig:containerisation} illustrates such a scenario: Service A runs inside a container with IP \texttt{172.17.0.2} on host \texttt{192.168.2.1} and initiates a connection to Service B, which also resides in a container with IP \texttt{172.17.0.2} on a different host, \texttt{192.168.2.2}, where port 80 is exposed (e.g., via Docker’s \texttt{-p 80:80} flag). Notably, because containers are only routable from within their respective hosts, each host can assign the same IP range to its container bridge network. As a result, both containers can have identical IP addresses without conflict, and the packet is still routed correctly across hosts.

Two modifications occur to packet's metadata in Figure~\ref{fig:containerisation}. First, as the packet exits host \texttt{192.168.2.1}, SNAT replaces packet's source IP with host's IP to ensure response routing. Second, upon arriving at host \texttt{192.168.2.2}, DNAT redirects the packet to container IP due to the exposed port $80$. These NAT rules reverse on the return path, i.e. DNAT becomes SNAT and SNAT becomes DNAT.

Above example configuration demonstrates the fundamental challenge of inferring true communicating service identities solely from IP/TCP metadata. No observation point in Figure~\ref{fig:containerisation} reveals the true interacting entities, i.e., Services A and B running on hosts \texttt{192.168.2.1} and \texttt{192.168.2.2}. One mitigation is to augment packet metadata with awareness of NAT state on the hosts performing translation. Coroot~\cite{coroot} implements such approach, inspects each host’s \texttt{conntrack} tables to resolve the packet’s ultimate destination. However, as shown next, Coroot’s approach fails when NAT occurs outside administratively controlled hosts.
% As illustrated by the various observation points in Figure~\ref{fig:containerisation}, there is no point of observation that uncovers the true interacting entities, i.e., Services A and B running on hosts \texttt{192.168.2.1} and \texttt{192.168.2.2}. One way to mitigate this limitation is to augment packet metadata with awareness of NAT state on the hosts performing translation. This is the approach taken by Coroot~\cite{coroot}, which inspects each host’s \texttt{conntrack} table to resolve the packet’s ultimate destination. However, as the next section will show, even this method falls short when NAT is performed by infrastructure components beyond our administrative control.

\subsection{Cross-Network Connectivity via NAT}
\label{sec:external-nat}
{\color{black}
To mitigate the exhaustion of the IPv4 address space, NAT enables multiple devices within a private LAN to share a single public IP address~\cite{tsuchiya1991ip,olteanu2015lost}. This form of address translation is ubiquitous in modern network infrastructures, as it provides standard Internet connectivity to privately addressed hosts. A common instance of this configuration occurs in cloud environments: when communicating with a virtual machine (VM) using its public IP address, a network device external to the VM performs DNAT, rewriting the packet’s destination to the VM’s private IP address. This scenario resembles the container-based setup in Section~\ref{sec:containerisation}, with a key difference that NAT is performed by an external device rather than by the host itself. 

% To mitigate the exhaustion of the IPv4 address space, Network Address Translation (NAT) enables multiple devices within a private LAN to share a single public IP address~\cite{tsuchiya1991ip,olteanu2015lost}. This form of address translation is ubiquitous in modern network infrastructures, as it provides standard Internet connectivity to privately addressed hosts. A common instance of this configuration arises in cloud environments: when communicating with a virtual machine (VM) via its public IP address, a network device external to the VM performs destination NAT (DNAT), rewriting the packet’s destination to the VM’s private IP address.
} 
% This scenario resembles the container-based setup in Section~\ref{sec:containerisation}, but differs in that an external device performs the NAT rather than the host processing the packet.

\begin{figure}[t]
\centering
\upp
\includegraphics[width=\columnwidth]{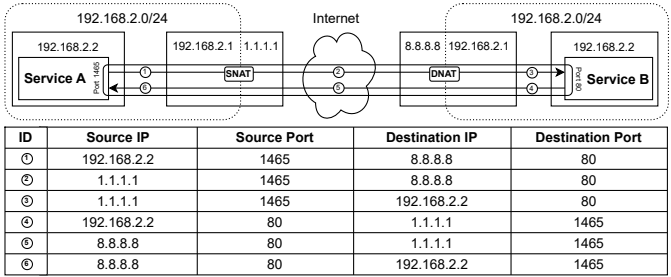}
\upp\upp
%\upp
\caption{\textbf{External NAT network configuration.} Services A  and B run on hosts in separate LANs, with NAT applied at each LAN's boundary.}\upp
\label{fig:cross-network} 
\upp
%\upp
%\upp\upp\upp
\end{figure}

\sloppy
Figure~\ref{fig:cross-network} depicts this configuration where a router is responsible for providing connectivity between devices on different LANs. Service A runs on a host within a LAN using private IPv4 range \texttt{192.168.2.0/24}, connected to a router with private IP \texttt{192.168.2.1} and public IP \texttt{1.1.1.1}. Service B operates in a separate LAN that uses the same private address range behind a router configured with private IP \texttt{192.168.2.1} and public IP \texttt{8.8.8.8}. Similar to previous setup (see Section~\ref{sec:containerisation}), the packet transforms twice. First, SNAT changes the source IP to the sending router's public IP (\texttt{1.1.1.1}) as it exits the LAN. Second, the receiving router applies a DNAT rule to translate destination IP to Service B's host internal address upon reaching destination network. Since both networks use private address spaces, overlapping IP ranges cause no routing issues across the public Internet.

As in the container-based setup discussed in Section~\ref{sec:containerisation}, inspecting a packet’s source and destination addresses alone (e.g., Caretta~\cite{caretta}) proves insufficient for accurately identifying true communicating endpoints. From Service A's perspective, the communication incorrectly appears to occur with Gateway 2, while Service B perceives the remote host as Gateway 1. However, unlike the previous setup in Section~\ref{sec:containerisation}, the gateways in this scenario typically lie outside the domain of administrative control. Consequently, techniques relying on host-level NAT state (e.g., Coroot~\cite{coroot}) cannot access the necessary translation tables and thus fail to reconstruct the connection's true endpoints. Such approaches yield incomplete and misleading results regarding the actual communicating services, inconsistencies as demonstrated experimentally in 
Section~\ref{sec:results-reconstruction}. The following section introduces Ripple’s design and methodology to address these shortcomings.

\section{Design}
Our goal is to develop an approach that discovers a distributed system's service dependency graph despite the adversities presented in Section~\ref{sec:problem-formulation}. Observer approaches that rely on a packet's metadata to infer communicating services cannot see through NAT, which leads to inaccurate and undiscovered relationships. Motivated by these challenges, we develop Ripple, a run-time modifier instrumentation tool, which aims to meet the following design goals:
\begin{enumerate}[label=\textcircled{\scriptsize\bfseries\arabic*}]
    \item \label{objective:protocol} \textbf{Protocol Agnosticism.} \textit{Support all application-layer (layer 5+) protocols built on top of TCP, including encrypted protocols} - This objective broadens Ripple's discovery capabilities to userspace plaintext and encrypted protocols (e.g. HTTPS, gRPC). This ensures applicability in encryption-standard modern production environments.
    %modern production environments where encryption is the norm.
    \item \label{objective:granularity} \textbf{Fine-Grained Visibility.} \textit{Capture service dependencies at a host$\rightarrow$process level of granularity} - Existing approaches (e.g. Caretta~\cite{caretta}) settle for identifying the communicating network hosts, which is sufficient in containerised setups where each container typically runs a single service. However, in traditional multi-process environments, such as bare-metal deployments or VMs, many services share a namespace, making it essential to disambiguate connections at the process level.
    \item \label{objective:userspace-processes} \textbf{Userspace Dependency Mapping.} \textit{Build the service dependency graph based on all userspace processes that communicate with each other} - By anchoring the graph in userspace processes, Ripple can expose not just the end services, but also any intermediary actors such as layer 7 load balancers or proxies \cite{shi2025miresga, gandhi2016yoda} (e.g., Envoy \cite{envoy}, HAProxy \cite{haproxy}, NGINX \cite{nginx}) that route or modify traffic. These are services that can have a real impact on the experienced performance by an end-user, and therefore, must be included in the resulting service dependency map.
    \item \label{objective:graceful-deployment} \textbf{Graceful Deployment.} \textit{Avoid breaking existing communication even when Ripple is deployed on only a subset of hosts} - This ensures that Ripple can be incrementally adopted in production environments without requiring full network coverage or risking service disruption, thereby lowering the barrier to adoption.
    \item \label{objective:non-intrusive} \textbf{Non-Intrusiveness.} \textit{Operate without requiring source code instrumentation} - This enables Ripple to work with unmodified binaries and third-party applications, removing the burden of application-level integration and simplifying deployment.
\end{enumerate}

The key insight underpinning Ripple recognizes that passively observing metadata from IP and TCP headers often fails to accurately identify true connection endpoints, since intermediate nodes (e.g., NAT devices) modify this information during packet transit. Instead, Ripple leverages that a process injecting data into an existing connection routes it to the same recipient as the original communication, unaffected by address translations. Because injected data does not alter the routing path, this reliably attributes the connection to the actual userspace process receiving the traffic.
% Instead, Ripple leverages the fact that a process able to inject data into an existing connection will route that data to the same recipient as the original communication, regardless of intervening address translations. Because this injected data does not influence the routing path, it reliably attributes the connection to the actual userspace process receiving the traffic.

However, naively injecting a fabricated packet into an active TCP connection without accounting for TCP’s state machine risks violating design objective \ref{objective:graceful-deployment}, particularly when no agent is present on the receiving end to discard unexpected data. In this context, an agent refers to lightweight run-time components that enable service dependency discovery.

To avoid state violations, both IPv4 and TCP options present potential mechanisms for embedding additional context directly into packets while remaining independent of application-layer visibility and compatible with encrypted traffic. However, prior work has shown that the IPv4 options field is poorly supported~\cite{gont2014rfc,fransson2004end} across network devices, often resulting in increased delay, jitter, and packet loss. In contrast, due to the increased reliability provided by TCP options~\cite{honda2011still}, Ripple opted for these to embed metadata without interfering with a connection’s state or data stream, while ensuring the metadata routes identically to application payload.

Consequently, the task of identifying communicating processes reduces to ensuring an agent is present at both endpoints and attributing the metadata to sending and receiving processes, thereby addressing objectives \ref{objective:protocol}, \ref{objective:granularity}, and \ref{objective:userspace-processes}. Crucially, because the experimental option resides in the TCP header rather than the payload, the connection remains undisturbed even without a receiving agent, preserving compliance with design objective \ref{objective:graceful-deployment}. As a result, Ripple supports both encrypted and non-encrypted traffic, without modifying or inspecting the original payload. Finally, Ripple achieves objective \ref{objective:non-intrusive} by injecting Traffic Control \emph{eBPF} programs (see details in Section~\ref{sec:implementation}) at run-time, which enables metadata insertion without requiring changes to application code or binaries.

To summarize, a Ripple agent deploys on each operating system hosting Services A and B. When a Ripple-instrumented process initiates communication through an established TCP connection, agent intercepts the outgoing packet on egress path, appends a \verb|16B unique identifier| to the TCP header’s \texttt{Options} field, and reinserts the packet into the operating system’s networking stack for normal transmission. Upon arrival at receiving host, the local Ripple agent intercepts the packet on the ingress path and monitors the system to determine which process consumes the packet’s payload. To clarify Ripple’s operation, the next section describes the core components and techniques that realise its design goals.

% To clarify Ripple’s practical operation, following section explores its implementation by describing the core components and techniques that realise its design goals and objectives.

\section{Implementation}
\label{sec:implementation}
\begin{figure*}[t]
\centering  
\upp
\includegraphics[width=\textwidth]{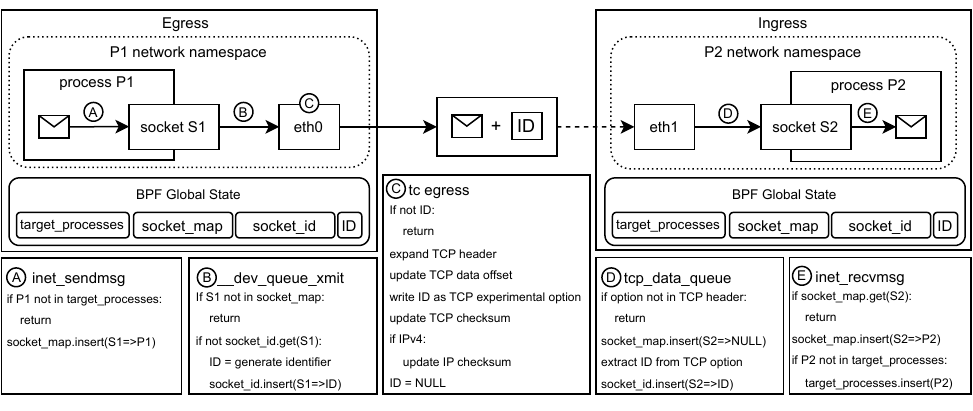}
\upp\upp\upp
%\upp\upp
\caption{\textbf{Overview of Ripple’s discovery process.} A packet transmitted through a connection associated with socket \texttt{S1}, owned by a \textit{target} process \texttt{P1}, triggers the execution of the egress \textit{eBPF} programs \texttt{A} and \texttt{B}, which determine whether the packet requires instrumentation. If so, program \texttt{C} injects an identifier \texttt{ID} into the packet and returns it to the networking stack for normal processing. On ingress, program \texttt{D} detects the identifier and records the receiving socket \texttt{S2}, while program \texttt{E} links the socket to the receiving process \texttt{P2}, thereby completing service discovery for that connection.}
\upp\upp
\label{fig:architecture}
\end{figure*}
Rather than instrumenting all processes running on host, Ripple analyses only a list of target processes—those either explicitly specified or dynamically discovered. A process joins this target set in one of two ways:
\begin{enumerate*}[label=(\roman*)]
\item when it receives a packet containing a Ripple-injected identifier, or
\item through manual registration, which allows Ripple to bootstrap discovery from a known process and incrementally expand service dependency graph as new dependencies emerge.
\end{enumerate*}

{\color{black}
Moreover, because Ripple operates over the stateful TCP protocol, it injects its identifier \textbf{only once} per connection; into the first data packet whose size does not exceed $MTU$ - \SI{16}{\byte}. This single insertion is sufficient to establish a persistent process-process association that is maintained for the entire lifetime of the connection, even if the network path changes. This property follows directly from TCP’s connection-oriented semantics and also explains why the approach is not extended to UDP, whose stateless design would require repeated instrumentation. By instrumenting only a target set of processes and avoiding per-packet modification, Ripple significantly reduces instrumentation frequency and overhead, improving scalability.

% By concentrating exclusively on a target set of processes and avoiding per-packet modification, this design significantly reduces instrumentation frequency and overhead, thereby improving scalability.
% Moreover, because Ripple operates over the stateful TCP protocol, it injects its identifier \textbf{only once} per connection; into the first data packet whose size does not exceed $MTU$ - \SI{16}{\byte}
% %- 16,\text{B}$. 
% This single insertion is sufficient to establish a persistent process to process association that is maintained for the entire lifetime of the connection, even in the presence of network path changes. This property follows directly from TCP’s connection-oriented semantics and also explains why the approach is not extended to UDP, whose stateless design would require repeated instrumentation. By concentrating exclusively on a target set of processes and avoiding per-packet modification, this design significantly reduces instrumentation frequency and overhead, thereby improving scalability.
}

Figure~\ref{fig:architecture} presents an overview of Ripple’s custom instrumentation programs and their corresponding pseudocode for discovering communicating services within a distributed system. When a target process writes data to a socket, it generates a packet and passes it to a network interface for routing to its destination. This sequence forms the packet’s egress path and triggers Ripple’s \emph{eBPF-based instrumentation programs} \circled{A}, \circled{B}, and \circled{C}. Programs \circled{A} and \circled{B} track the process initiating data transfer and determine whether the outgoing packet originates from a target process and requires tagging an identifier. If so, program \circled{C} injects the identifier into the packet.

On the receiving end, when a packet enqueues on a socket, program \circled{D} inspects it for a Ripple-injected identifier. If present, the program records the socket in a global map. Later, when a process reads from the socket, program \circled{E} links it to the recorded socket and adds it to the target process list. This mechanism enables discovery to propagate across both existing and future connections. Remainder of this section details Ripple’s egress and ingress components.

\subsection{Egress}
\label{sec:egress}
% Ripple's egress instrumentation is responsible for injecting identifiers into outbound packets originating from target processes. These identifiers allow Ripple agents on the receiving end to associate incoming traffic with its originating process, thereby enabling process-level dependency mapping.

Ripple’s egress instrumentation injects identifiers into outbound packets originating from target processes, letting receiving agents associate incoming traffic with its originating process and enabling process-level dependency mapping.

To determine whether an outbound packet (\texttt{struct sk\_buff *skb}) should be instrumented, Ripple initially inspects the originating socket and the process owning that socket. If the socket is associated with one of Ripple's target processes, and no discovery identifier has yet been sent for that connection (socket), Ripple marks the packet for instrumentation. The latter condition ensures that only the first packet of each connection is used for dependency discovery, thereby significantly reducing overhead and avoiding redundant annotations for long-lived connections.

Ripple tracks socket ownership using \emph{eBPF} kernel probes attached to \texttt{inet\_sendmsg} and \texttt{inet6\_sendmsg}. These hooks capture the socket descriptors from target processes and store them in a global \emph{eBPF} hash map. Determining whether a target process generated a packet is then reduced to checking whether the creating socket is recorded in this map. Ripple enables this by instrumenting the \texttt{\_\_dev\_queue\_xmit} kernel function (program \texttt{B}), which marks the point where a packet’s ownership transfers from the socket to its first network interface. At this stage, program \texttt{B} extracts the \texttt{skb->sk->sk\_socket} pointer and performs the originating socket check; if the socket is registered in the global \emph{eBPF} hash map, and provided Ripple has not yet emitted a discovery event for that socket, program \texttt{B} marks the packet for the injection by the final egress program \texttt{C}.

Injecting an identifier in the packet is done by an \emph{eBPF} TC egress program (program \texttt{C}) attached to all relevant network interfaces. When this program encounters a packet that meets the conditions for identifier injection, it performs the following steps:
\begin{enumerate*}[label=(\roman*)]
\item Reserves space in the TCP header to accommodate a custom TCP option that will carry the identifier;
\item Updates the data offset field to reflect the new size of the header;
\item Writes the identifier into the allocated TCP option space;
\item Recalculates the TCP checksum to account for the modified header;
\item If the packet uses IPv4, updates the IP header checksum to reflect the changes; for IPv6, this step is unnecessary due to differences in protocol semantics; and
\item Re-injects the modified packet into the networking stack for standard routing and delivery.
\end{enumerate*}

By intercepting and modifying the packet at this level, Ripple ensures the injected metadata is delivered along the same routing path as regular application data, without disrupting the connection or requiring changes to the application itself. This egress instrumentation is crucial for maintaining compliance with Ripple’s design goals (Objectives \ref{objective:protocol} - \ref{objective:non-intrusive}).

\subsection{Ingress}

\begin{figure*}[t]
    \centering
    \subfloat[\label{fig:ripple-k8s-online}]{
      \includegraphics[width=0.33\textwidth]{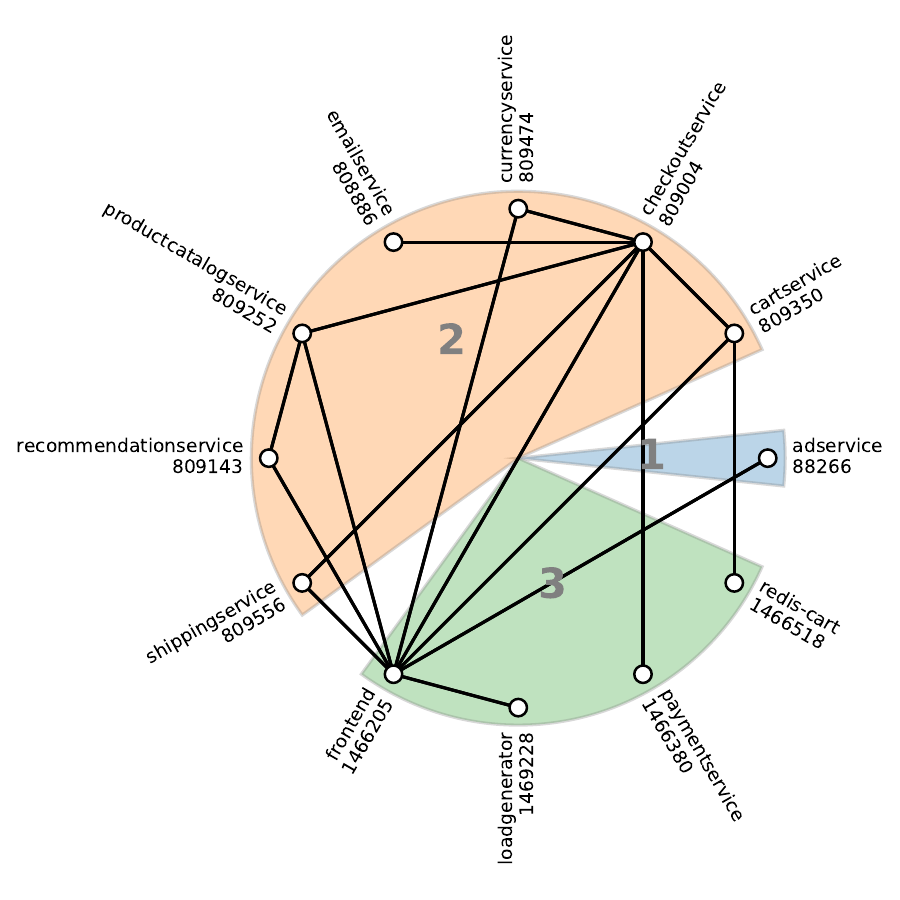}
    }
    \subfloat[\label{fig:ripple-k8s-media}]{
      \includegraphics[width=0.33\textwidth]{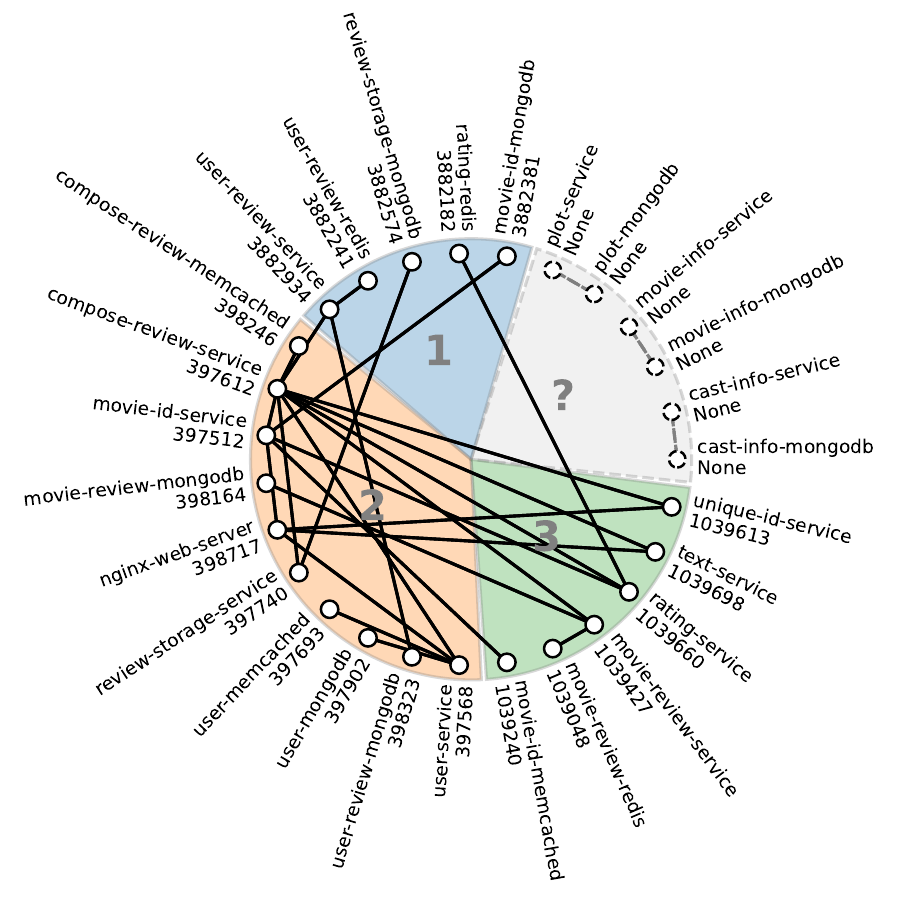}
    }
    \subfloat[\label{fig:ripple-k8s-social}]{
      \includegraphics[width=0.33\textwidth]{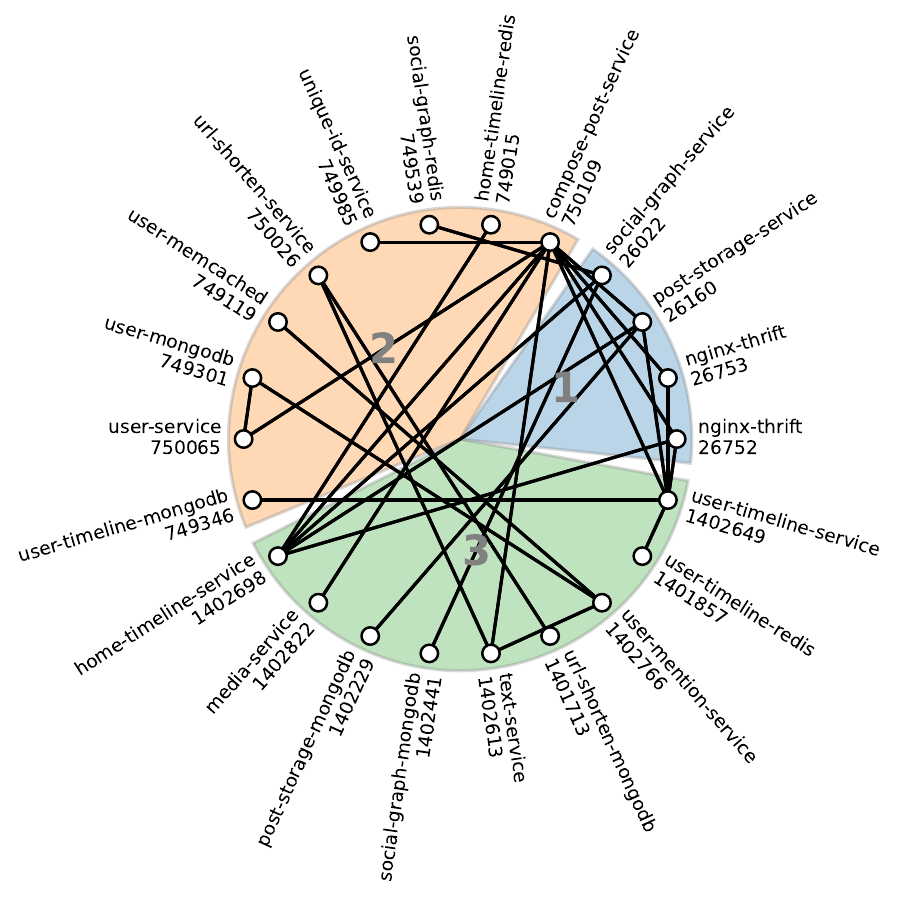}
    } 
    \upp
    %\upp
    \caption{Ripple service map reconstruction for the NAT-free network configuration. (a) Online Boutique~\cite{online-boutique}; (b) Media~\cite{gan2019open}; (c) Social Media~\cite{gan2019open}.}
    \label{fig:ripple-k8s-reconstruction} \upp
    %\upp
\end{figure*}

On packet ingress, Ripple addresses two challenges: \begin{enumerate*}[label=(\roman*)] \item detecting inbound TCP packets carrying a discovery identifier previously injected by a Ripple agent; and
\item determining userspace process ultimately receiving the packet to complete dependency mapping.
\end{enumerate*}

Ripple detects the injected identifier using an \emph{eBPF} probe attached to \texttt{tcp\_data\_queue} kernel function, which triggers when a packet queues onto a TCP socket. Next, it inspects TCP header of associated \texttt{struct sk\_buff *skb} for its custom TCP option. Upon finding the identifier, Ripple records corresponding \texttt{struct sock *sk}, passed as an argument to \texttt{tcp\_data\_queue}, in a global eBPF hash map for further resolution.

To complete identification of the receiving process, Ripple correlates the socket with its owning userspace process. It achieves this by attaching \emph{eBPF} probes to \texttt{inet\_recvmsg} and \texttt{inet6\_recvmsg} kernel functions, which are invoked when a process reads from a socket. When such an event occurs, Ripple checks whether the accessed socket matches one recorded during earlier identifier detection. If a match is found, Ripple links the process to socket that received the identifier, completing the discovery of receiving entity.

If the receiving process is not already in Ripple’s target process list, Ripple dynamically registers it. This registration attaches Ripple’s TC egress programs (Section~\ref{sec:egress}) to all network interfaces within the process’s namespace. Consequently, Ripple continues propagating discovery events from newly identified processes, enabling the incremental construction of an accurate service dependency graph without disrupting application activity.

\subsection{Metric Collection}
Ripple is implemented as a TCP discovery extension of the Prism~\cite{landau2026beyond,landau2026threadstatesdiagnosingperformance} metric collector. Prism is a Linux performance diagnostic tool for analysing thread dynamics in distributed systems. It collects lightweight eBPF-based metrics across scheduling, futexes, I/O, networking, and storage, enabling fine-grained analysis of application performance at the level of individual threads, their interactions, and their interactions with system resources. 

Ripple adds a new discovery subsystem, wherein the identifier is propagated between different agents. This module registers \emph{eBPF} Programs \circled{A}-\circled{E}, communicating the socket~$\rightarrow$~identifier mapping to the userspace metric collector via a \texttt{BPF\_MAP\_TYPE\_RINGBUF} \emph{eBPF} map type. Ripple also extends Prism's VFS subsystem by collecting the total amount of bytes exchanged by a process to a particular socket. These metrics are statistically summarised and collected at a \texttt{second} granularity, with logarithmic histograms capturing the more granular sub-second timings.

\begin{figure}[!t]
  \scriptsize
  \centering
  \upp
  %\upp
  %\upp\upp
  \begin{lstlisting}[
      caption={Templated query that computes the process dependency graph between period \texttt{t1} and \texttt{t2}},
      style=sqlstyle,
      label={lst:dependency-graph},
      frame=none,
      basicstyle=\ttfamily\footnotesize,
      xleftmargin=0pt,
      xrightmargin=0pt
  ]
WITH
  pid_sock AS (
    SELECT DISTINCT machine_id, pid, inode_id
    FROM vfs
    LEFT JOIN linux_consts lc ON vfs.fs_magic = lc.value AND lc.const_type = 'fs_magic'
    WHERE lc.const_name = 'SOCKFS_MAGIC'
      AND (ts_s >= {{ t1 }} AND ts_s <= {{ t2 }})
  ),
SELECT
  pid1.machine_id AS machine1, pid1.pid AS pid1, pid1.inode_id AS sock1,
  pid2.machine_id AS machine2, pid2.pid AS pid2, pid2.inode_id AS sock2
FROM discovery d
INNER JOIN pid_sock pid1 ON d.local_machine_id  = pid1.machine_id AND d.local_inode_id  = pid1.inode_id
INNER JOIN pid_sock pid2 ON d.remote_machine_id = pid2.machine_id AND d.remote_inode_id = pid2.inode_id
  \end{lstlisting} \upp\upp
  %\upp
\end{figure}

In terms of data storage, Prism offers two primary deployment configurations: a standalone setup, in which the collected telemetry data is stored in a \texttt{DuckDB} database~\cite{raasveldt2019duckdb}, and a distributed setup, in which the agents store their data in a sharded \texttt{clickhouse} database~\cite{schulze2024clickhouse}. Regardless of the selected data store, Listing~\ref{lst:dependency-graph} demonstrates how to construct the service dependency graph. The query first filters the sockets, and by extension the processes that actively exchanged data between $[\texttt{t1}; \texttt{t2}]$ via the \texttt{vfs}. The \texttt{discovery} table is then used to join the two processes that communicated with one another, completing the process dependency mapping for the period of interest. Prism also collects process metadata, such as container labels, which can be useful for adding more context to the information provided by Listing~\ref{lst:dependency-graph}. The results presented in Section~\ref{sec:results-reconstruction} leverage this metadata to construct the dependency graphs in Figure~\ref{fig:ripple-k8s-media}.
% With regards to data storage, Prism provides two main deployment configurations: A standalone setup which stores the collected telemetry data in a DuckDB database; and a distributed setup where the agents store their data in a sharded clickhouse database. Irrespective of the selected data store, Listing~\ref{lst:dependency-graph} illustrates how to construct the service dependency graph. The query first filters the sockets, and by extension the processes, that actively exchanged data between $[\texttt{t1}; \texttt{t2}]$ via the \texttt{vfs}. The \texttt{discovery} table is then used to join the two processes that communicated with one another, completing the process dependency mapping for the period of interest. Prism also collects process metadata, such as container labels, which can be useful to add more context to the information provided by Listing~\ref{lst:dependency-graph}. The results presented in Section~\ref{sec:results-reconstruction} leverage this metadata to construct the dependency graphs in Figure~\ref{fig:ripple-k8s-media}.

\section{Evaluation}
\label{sec:evaluation}

This section evaluates Ripple across three key dimensions: service dependency map reconstruction (Section \ref{sec:results-reconstruction}), run-time overhead (Section \ref{sec:results-overhead}), and scalability (Section \ref{sec:results-scalability}). To ensure relevance and comparability, we conduct our experiments using three widely adopted microservice benchmarks: the Social Media Microservice Benchmark \cite{gan2019open}, the Media Microservice \cite{gan2019open}, and Online Boutique \cite{online-boutique}. 
Previous works~\cite{xie2024pbscaler,ashok2024traceweaver,zhang2024mucache,saleh2022empirical,zhu2023dissecting} have extensively used these benchmarks, which offer a representative mix of service communication patterns, deployment topologies, programming languages, and traffic characteristics observed in real-world distributed systems.

Section \ref{sec:results-reconstruction} evaluates the correctness and completeness of Ripple’s reconstructed dependency graphs, using controlled experiments across multiple networking configurations, similar to those presented in Section \ref{sec:problem-formulation}. These experiments enabled us to demonstrate Ripple’s robustness in a variety of network topologies representative of real-world deployment scenarios. This section also compares Ripple to the state-of-the-art approaches (i.e., Caretta~\cite{caretta}, Coroot~\cite{coroot}, Deepflow~\cite{shen2023network}) introduced in Section~\ref{sec:related-works}, highlighting their inability to adapt to heterogeneous networking scenarios.

Section~\ref{sec:results-overhead} presents an overhead analysis that quantifies run-time impact of both Ripple and state-of-the-art instrumentation approaches (i.e., Caretta~\cite{caretta}, Coroot~\cite{coroot}, Deepflow~\cite{shen2023network}) on the monitored system, with a particular focus on the latency introduced in benchmarked microservice applications. 
To evaluate scalability, Section~\ref{sec:results-scalability} examines Ripple’s performance with increasing number of monitored processes and inter-service dependencies. To assess this, we use a \emph{time-to-completion} metric, which captures the duration required to reconstruct the architecture of the instrumented system.

Lastly, to demonstrate Ripple's practical utility in operational settings, we include a motivating example in Section~\ref{sec:results-motivating-example} that illustrates its role in root cause localisation. This example shows how Ripple can guide operators toward the relevant services responsible for performance degradation in a distributed system, thereby reducing time to resolution and narrowing the diagnostic scope.

Ripple is developed as a Rust executable that incorporates \emph{eBPF} programs through \emph{libbpf-rs} \cite{libbpf-rs}, and is therefore currently limited to Linux systems. Its complete source code and reproducibility artifact
% \footnote{Artifact is also included in the recommended two-page Artifact Description (AD)/Artifact Evaluation (AE) appendix, maintaining double-blind anonymity.} 
is available in an \emph{online} \verb|GitHub| repository~\cite{repo,repro}.
To evaluate Ripple, we conducted our experiments on a cluster of three Amazon Web Services (AWS) \verb|t2.large| virtual machines (VMs), with $2$ vCPUs and \SI{8}{\giga\byte} each, running Ubuntu \verb|24.04.2 LTS| with \verb|Linux 6.11.0| kernel. The cluster was initialized using \texttt{kubeadm} \cite{kubeadm} and runs Kubernetes version \verb|v1.33.0|. It uses \texttt{containerd} \cite{containerd} as the container run-time and \texttt{flannel} \cite{flannel} as the container network interface. For experiments requiring Docker, we use version \verb|28.1.1|. All VMs reside within the same private network and are reachable via both their internal (private) IP addresses and their externally assigned AWS-managed public IPs.

\subsection{Service Map Reconstruction}
\label{sec:results-reconstruction}

\begin{table*}[!t]
\centering
\caption{Service map reconstruction correctness across tools, benchmarks, network configurations using precision, recall, $\text{F}_1$ score.}
\label{table:reconstruction-results}
\vspace*{-0.1cm}
\upp
\resizebox{\linewidth}{!}
{
    \begin{tabular}{|l|l|r|r|r|r|r|r|r|r|r|}
    \hline
    &  & \multicolumn{3}{c|}{\bf Social} & \multicolumn{3}{c|}{\bf Media} & \multicolumn{3}{c|}{\bf Online-boutique} \\
    \cline{3-11}
   {\bf Approach} & {\bf Metrics} & {\bf NAT-free} & {\bf internal NAT} & {\bf external NAT} & {\bf NAT-free} & {\bf internal NAT} & {\bf external NAT} & {\bf NAT-free} & {\bf internal NAT} & {\bf external NAT} \\
    \hline
    \hline
    \multirow{3}{*}{Caretta} & precision & 1.00 & 0.00 & 0.00 & 1.00 & 0.00 & 0.00 & 1.00 & 0.00 & 0.00 \\ \cline{2-11}
    & recall & 1.00 & 0.00 & 0.00 & 0.79 & 0.00 & 0.00 & 1.00 & 0.00 & 0.00 \\ \cline{2-11}
    & $\text{F}_1$ & \cellcolor{bestgreen}1.00 & \cellcolor{badred}NaN & \cellcolor{badred}NaN & \cellcolor{bestgreen}0.89 & \cellcolor{badred}NaN & \cellcolor{badred}NaN & \cellcolor{bestgreen}1.00 & \cellcolor{badred}NaN & \cellcolor{badred}NaN \\ 
    \cline{1-11}
    \multirow{3}{*}{Coroot} & precision & 1.00 & 1.00 & NaN & 1.00 & 1.00 & 0.24 & 1.00 & 1.00 & 0.46 \\ \cline{2-11}
    & recall & 1.00 & 0.88 & 0.00 & 0.79 & 0.79 & 0.09 & 0.81 & 0.75 & 0.19 \\ \cline{2-11}
    & $\text{F}_1$ & \cellcolor{bestgreen}1.00 & \cellcolor{goodgreen}0.94 & \cellcolor{badred}NaN & \cellcolor{bestgreen}0.89 & \cellcolor{bestgreen}0.89 & \cellcolor{mediocre}0.13 & \cellcolor{goodgreen}0.90 & \cellcolor{goodgreen}0.86 & \cellcolor{mediocre}0.27 \\
    \cline{1-11}
    \multirow{3}{*}{Deepflow} & precision & 1.00 & NaN & NaN & 1.00 & NaN & NaN & 1.00 & NaN & NaN \\
    & recall & 0.52 & 0.00 & 0.00 & 0.53 & 0.00 & 0.00 & 1.00 & 0.00 & 0.00 \\ \cline{2-11}
    & $\text{F}_1$ & \cellcolor{mediocre}0.68 & \cellcolor{badred}NaN & \cellcolor{badred}NaN & \cellcolor{mediocre}0.69 & \cellcolor{badred}NaN & \cellcolor{badred}NaN & \cellcolor{bestgreen}1.00 & \cellcolor{badred}NaN & \cellcolor{badred}NaN \\
    \cline{1-11}
    \multirow{3}{*}{Ripple} & precision & 1.00 & 1.00 & 1.00 & 1.00 & 1.00 & 1.00 & 1.00 & 1.00 & 1.00 \\ \cline{2-11}
    & recall & 1.00 & 1.00 & 1.00 & 0.71 & 0.71 & 0.71 & 1.00 & 1.00 & 1.00 \\ \cline{2-11}
    & $\text{F}_1$ & \cellcolor{bestgreen}1.00 & \cellcolor{bestgreen}1.00 & \cellcolor{bestgreen}1.00 & \cellcolor{goodgreen}0.83 & \cellcolor{goodgreen}0.83 & \cellcolor{bestgreen}0.83 & \cellcolor{bestgreen}1.00 & \cellcolor{bestgreen}1.00 & \cellcolor{bestgreen}1.00 \\
    \cline{1-11}
    \end{tabular}
}
\end{table*}

The goal of this experiment is to evaluate Ripple against state-of-the-art approaches (i.e., Caretta~\cite{caretta}, Coroot~\cite{coroot}, Deepflow~\cite{shen2023network}) in terms of their ability to accurately reconstruct service dependency graphs, measured using \emph{precision} (correctly identified dependencies) and \emph{recall} (completeness of discovered dependencies) metrics. For evaluation, we used three widely adopted microservice benchmarks: Online Boutique \cite{online-boutique} (12 services, 16 dependencies); Social Media Microservices \cite{gan2019open} (21 services, 25 dependencies); and Media Microservices \cite{gan2019open} (29 services, 34 dependencies), similar to related works~\cite{xie2024pbscaler,ashok2024traceweaver,zhang2024mucache,saleh2022empirical,zhu2023dissecting}.

To assess the robustness of each approach we evaluate them over each benchmark application under three different network configurations. The first configuration deploys all services within a single Kubernetes cluster configured with \texttt{flannel} as Container Network Interface (CNI) plugin. In this setup, inter-service communication occurs without NAT, similar to NAT-free topology in Section \ref{sec:nonat}.

The second configuration splits each microservice benchmark into $3$ separate \verb|docker compose| deployments and binds each service to a port exposed on its respective VM. Furthermore, when one service communicates with another, it uses the private IP address of the VM where the target service is deployed. Each container deployed on a VM receives its own network namespace, so when a packet arrives at a host port exposed for a container, the system performs NAT to modify the packet's destination to the service in the container's network namespace (Section \ref{sec:containerisation}).

The third configuration is similar to the second, except that services communicate via the VMs’ public IPs. This change introduces an additional NAT layer, as incoming traffic is routed through an AWS-managed gateway, which maps the public IP to the appropriate VM's private IP. This external NAT scenario reflects the configuration discussed in Section \ref{sec:external-nat}. In total, each approach (i.e., Ripple, Caretta~\cite{caretta}, Coroot~\cite{coroot}, and Deepflow~\cite{shen2023network}) are evaluated across nine environments (three applications $\times$ three network topologies) providing a comprehensive view of their reliability and adaptability to real-world deployment conditions.

Table \ref{table:reconstruction-results} presents the evaluation results for each approach across the nine experimental configurations. For every combination of approach, microservice benchmark, and network topology, we report the resulting precision, recall, and F\textsubscript{1}, where F\textsubscript{1} is calculated as $\text{F}_1 = 2 \frac{\text{precision} \cdot \text{recall}}{\text{precision} + \text{recall}}$. In the following, we examine the performance of each approach with respect to different networking configurations and their reconstruction results across three different benchmarks.

In the NAT-free configuration, Coroot, Caretta, and Ripple exhibit similar and consistently strong performance. Each approach correctly identify the service dependencies and recover nearly all of the expected edges in each benchmark. However, for the Media and Social Media microservice benchmarks, none of the tools are able to reconstruct the full set of expected dependencies. This discrepancy stems from the fact that the ground truth topology reflects all potential service relationships defined in the application architecture, whereas the evaluated tools rely solely on observed run-time communication. As such, if two services do not exchange traffic during the experiment window, their relationship remains undiscovered.

Specifically, regarding the Media Microservice benchmark, Ripple achieves an F\textsubscript{1} score of $0.83$, slightly trailing behind Coroot and Caretta, which both report scores of $0.89$. This difference is attributable to Ripple missing three dependency edges that both Caretta and Coroot were able to identify. Figure~\ref{fig:ripple-k8s-media} visualizes the dependency graph discovered by Ripple, with the missing dependencies captured by Coroot and Caretta represented as dashed lines. A closer analysis reveals that this limitation is linked to Ripple’s overhead reduction strategy, which confines discovery events to originate only from processes that are part of its target set (see details in Section \ref{sec:implementation}). In this experiment, Ripple was bootstrapped on the \texttt{text-service} process, which resides in the main service graph but lacks connectivity to the three independent graphs related to the \texttt{cast}, \texttt{movie} and \texttt{plot} services. Because no packets are emitted to those disconnected components from a target process, Ripple never observes them, and therefore cannot report their dependencies. 

Figure~\ref{fig:ripple-k8s-reconstruction} illustrates fully resolved dependency graphs produced by Ripple for tested microservice benchmarks in NAT-free configuration. Each node corresponds to a process and is labeled with its \texttt{cgroup} and process identifiers, providing fine-grained visibility into the distributed system's structure. Nodes are clustered by machine identifier, clearly delineating cross-node communication and co-located services. Notably, process-level granularity targeted by Objective~\ref{objective:granularity} proves valuable in Social Media benchmark (Figure~\ref{fig:ripple-k8s-social}), where \texttt{nginx-thrift} service is shown to consist of two parallel processes sharing the load as front-end balancers, a detail that would be obscured by coarser-grained methods.

Deepflow, on the other hand, performs well on the Online Boutique benchmark but shows significantly lower recall on the Media and Social Media benchmarks, leading it to underperform in these scenarios. This degradation is attributed to architectural limitations noted by Yang et al. \cite{yang2025zerotracer}, who observe that Deepflow’s design is best suited to single-threaded applications and particular programming languages - performs well with PHP and Golang but not so well with Java and Python. As the Media and Social Media benchmarks exhibit more complex and concurrent communication patterns, Deepflow struggles to capture the full set of dependencies.

When switching to the internal NAT configuration, both Caretta and DeepFlow encounter limitations, as their Kubernetes-native designs prevent them from resolving container-to-container communication that traverses NAT. Instead, they attribute dependencies as occurring between the source container and the destination VM hosting the target service. This abstraction omits the true destination container, making both approaches unsuitable for environments that involve any form of NAT. 

In contrast, Coroot is able to reconstruct the service dependency map under this configuration. This is due to its ability to resolve NAT on a host by consulting \texttt{conntrack} table, allowing it to correctly trace communication to its actual container-level destination. Ripple similarly handles this configuration correctly, but with an important distinction that stems from its design goal outlined in Objective \ref{objective:userspace-processes}.

In Docker's default networking setup, services running on the same VM and communicating through a host's exposed port does not trigger the DNAT rule that is responsible for routing the data to the target container. Listing \ref{lst:docker-proxy} illustrates the iptables DNAT rule and \texttt{docker-proxy} service setup by Docker. The DNAT rule applies only when the incoming interface differs from the bridge interface to which the containers’ veth pairs are attached. As such, a packet originating from a container on this host trying to reach another container on the same host using the VMs private IP, will be forwarded through the L7 docker proxy service running in the host's network namespace.

\begin{figure}[!t]
\footnotesize,
\centering
\upp
\begin{minipage}{\linewidth} \upp\upp
\begin{lstlisting}[style=terminal, caption={Docker-related iptables rules, docker-proxy processes}, label={lst:docker-proxy}]
$ sudo iptables -t nat -L -n
Chain PREROUTING (policy ACCEPT)
 target prot opt in               out source     destination
 DOCKER 0    --  *                *   0.0.0.0/0  0.0.0.0/0   ADDRTYPE match dst-type LOCAL
Chain DOCKER (2 references)
 target prot opt in               out source     destination
 DNAT   6    --  !br-0cf265a146ac *   0.0.0.0/0  0.0.0.0/0   tcp dpt:5000 to:172.21.0.6:8080

$ ps -eo pid,cmd | grep docker-proxy
30709 docker-proxy -proto tcp -host-ip 0.0.0.0 -host-port 5000 -container-ip 172.21.0.6 -container-port 8080
30722 docker-proxy -proto tcp -host-ip :: -host-port 5000 -container-ip 172.21.0.6 -container-port 8080
\end{lstlisting}
\end{minipage}  \upp\upp\upp
\end{figure}

\begin{figure}[t]
    \centering
    %\upp
    \subfloat[\label{fig:docker-proxy-external}]{
      \includegraphics[width=0.51\columnwidth]{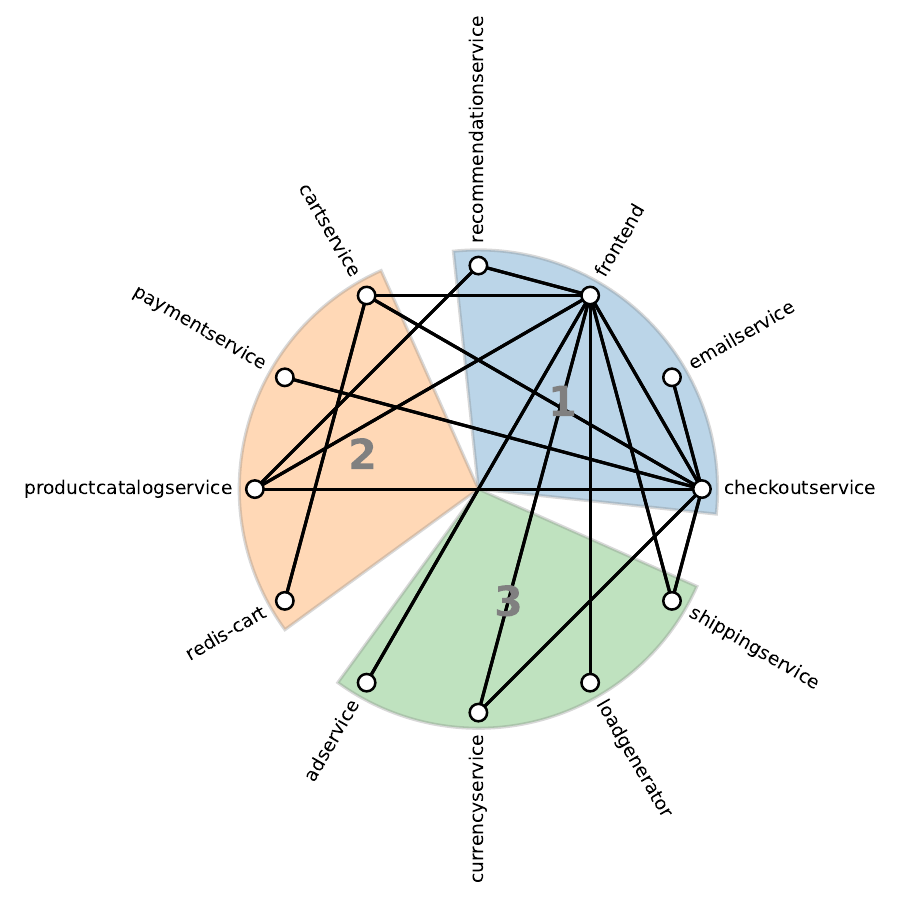}
    }
    \subfloat[\label{fig:docker-proxy-internal}]{
      \includegraphics[width=0.49\columnwidth]{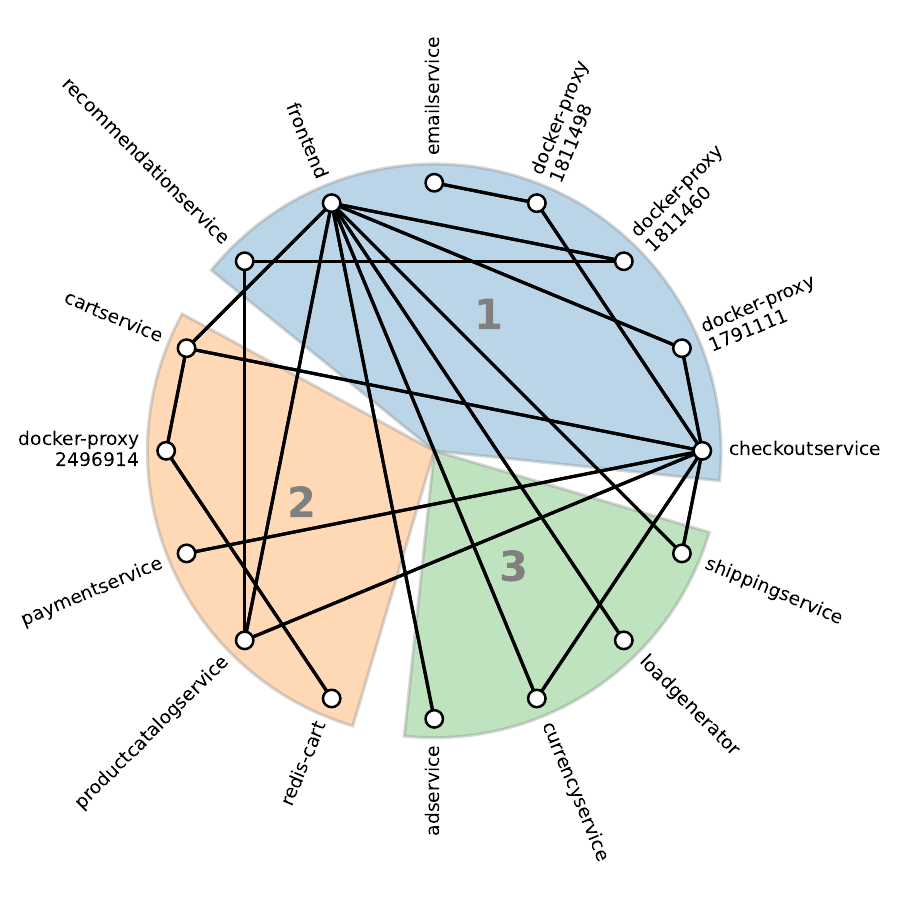}
    }\upp
    \caption{Online Boutique dependency graph in the internal NAT network configuration. (a) Coroot; (b) Ripple.}
    \label{fig:docker-proxy} \upp
\end{figure}

Unlike Coroot, Ripple captures this \texttt{docker-proxy} userspace process in its dependency graph, identifying it as an intermediary in the service communication path. Figure~\ref{fig:docker-proxy} illustrates Online Boutique's dependency graph as discovered by Coroot (figure~\ref{fig:docker-proxy-external}) and Ripple(figure~\ref{fig:docker-proxy-internal}). Notably, co-located services communicating with one another are directly connected in Coroot's dependency graph. However, as can be seen in Listing~\ref{lst:docker-proxy}, the data sent between these services is in fact forwarded by a \texttt{docker-proxy} service. This dependency is captured by Ripple, demonstrating its compliance with design goal~\ref{objective:userspace-processes}. Nevertheless, given that abstracting the \texttt{docker-proxy} processes makes the graphs equivalent, Table~\ref{table:reconstruction-results} considers both reconstructions as correct, despite Ripple offering additional visibility.

Lastly, in the external NAT configuration, both Caretta and DeepFlow are unable to reconstruct any service dependencies due to NAT. Coroot, which previously performed well, also fails under these conditions, since NAT now occurs outside the host running Coroot’s agents. This leads to Coroot’s inability to resolve the connection endpoints correctly, resulting not only in incomplete service dependency graphs but also in incorrect dependencies, with precision scores dropping to $0.13$ and $0.27$ for Media and Online Boutique benchmarks, respectively. In contrast, Ripple consistently reconstructs the correct service dependency graph across all benchmarks and network configurations. This underscores Ripple’s robustness and adaptability to a wide range of deployment environments, making it the only evaluated tool capable of maintaining high accuracy regardless of the underlying network topology.

\subsection{Overhead}
To assess the run-time overhead for each approach, we measure their impact on the request latency of the microservice benchmarks introduced in Section \ref{sec:results-reconstruction}. Each benchmark is deployed to our Kubernetes cluster, after which we initiate its standard load generator and record the average request latency over the duration of the experiment. For each approach-benchmark combination, we perform $30$ independent runs and report both the mean latency and standard deviation in Table \ref{table:overhead}. Baseline, refers to the scenario where the latency is measured without any of the instrumentation approaches.\label{sec:results-overhead}

Results show that Coroot introduces the least overhead, consistently yielding both the lowest latency increases and the most stable measurements across runs. Caretta and Ripple perform similarly, with latency overheads ranging from \SIrange[range-phrase=\,--\,]{0.7}{4.4}{\milli\second}, and standard deviations reported alongside the means. In Ripple’s case, the added latency stems primarily from the \emph{eBPF} programs shown in Figure~\ref{fig:architecture}, which execute additional code at each probe point. {\color{black} Moreover across all experiments in Section~\ref{sec:evaluation}, Ripple presented a maximum CPU usage of $\approx$ 10-12\% of a single core and sustained disk write rates of \SI{115}{\kilo\byte\per\second}.} This overhead could be further reduced by optimizing early return conditions and minimizing executed code paths, which we leave for future work. DeepFlow, by contrast, incurs the highest latency across all benchmarks, with average increases between \SIrange[range-phrase=\,--\,]{3.0}{8.3}{\milli\second}. While such overhead may be tolerable in many scenarios, it could pose challenges for latency-sensitive applications.

\begin{table}[t]
\centering
%\upp
\caption{Request latency average and standardard deviation for each approach-benchmark combination for a total of $30$ runs.} \upp
\label{table:overhead}
%\upp
%\setlength{\tabcolsep}{0.6pt}
\resizebox{\columnwidth}{!}
{
    \begin{tabular}{|l|r|r|r|}
    %\toprule
    \hline
   {\bf Approach} & {\bf Media} (\SI{}{\milli\second}) & {\bf Boutique} (\SI{}{\milli\second}) & {\bf Social} (\SI{}{\milli\second}) \\
    \hline
    \hline
    Baseline & 27.66 ± 1.27 & 40.28 ± 3.07 & 31.33 ± 2.03 \\ \hline
    Ripple & 28.35 ± 3.31 & 43.41 ± 4.52 & 33.96 ± 8.55 \\ \hline
    Caretta & 28.58 ± 1.69 & 44.66 ± 4.66 & 32.82 ± 9.27 \\ \hline
    Coroot & 28.71 ± 2.07 & 40.52 ± 3.63 & 32.99 ± 1.67 \\ \hline
    Deepflow & 35.93 ± 7.36 & 47.00 ± 4.50 & 34.33 ± 7.47 \\
    \hline
   % \bottomrule
    \end{tabular} 
    } \upp
\end{table}

\subsection{Scalability}
\label{sec:results-scalability}
As distributed systems grow in complexity, evaluating how Ripple scales when instrumenting the interconnected services that comprise an application becomes increasingly important. A key concern is \emph{time-to-completion}: the time Ripple takes to discover the full service dependency graph from the moment one of its agents is bootstrapped with a target process.
% As distributed systems grow in complexity, it becomes increasingly important to evaluate how Ripple scales when instrumenting the interconnected services that comprise an application. One key concern is the \emph{time-to-completion}, that is, the time Ripple takes to discover the full service dependency graph from the moment one of its agents is bootstrapped with a target process.

\begin{figure}[t]
\centering
%\upp
\includegraphics[width=0.35\textwidth]{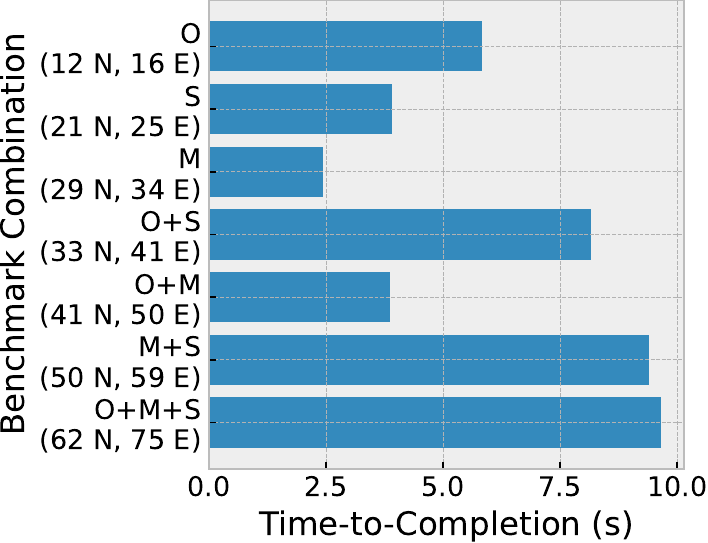}
%\upp
\caption{Time to complete service dependency discovery with Ripple, measured across individual benchmarks and in combined multi-benchmark deployments.}
\label{fig:ttc} \upp
\end{figure}

To evaluate this, we deploy the three microservice benchmarks from Section \ref{sec:results-reconstruction}, both in isolation and in various combinations. This allows us to systematically increase the number of services and inter-service dependencies to assess how these factors affect Ripple’s performance.

% entries = [
% ("O\n(12 N, 16 E)", 5.83),
% ("S\n(21 N, 25 E)", 3.91),
% ("M\n(29 N, 34 E)", 2.43),
% ("O+S\n(33 N, 41 E)", 8.16),
% ("O+M\n(41 N, 50 E)", 3.86),
% ("M+S\n(50 N, 59 E)", 9.40),
% ("O+M+S\n(62 N, 75 E)", 9.66),
% ]

Figure \ref{fig:ttc} presents a bar chart summarizing the time Ripple requires to fully reconstruct the service dependency graph across various benchmark configurations. Each bar corresponds to a combination of benchmarks, with the x-axis indicating the time-to-completion and the y-axis listing the benchmarks annotated with their respective number of Nodes (N) and Edges (E). For readability, we abbreviate the benchmark names as follows: O (Online Boutique), M (Media Microservices), and S (Social Media Microservices).

Interestingly, while the \emph{time-to-completion} generally increases with the overall number of services and dependencies, the trend is not strictly linear. For example, the Online Boutique + Media configuration (41 services, 50 edges) completes faster than the Online Boutique + Social Media setup (33 services, 41 edges), despite having more elements to discover. This suggests that Ripple’s completion time is influenced more by run-time communication patterns, i.e., when and how services exchange traffic, than by the sheer number of services or edges alone.

Nevertheless, as the total number of services increases, the likelihood of encountering services that delay discovery due to infrequent or delayed communication also rises. This leads to greater variation in completion time across larger configurations.

\subsection{Motivating Use Case}
\label{sec:results-motivating-example}

One of Ripple’s use case is narrowing down set of services responsible for observed performance degradation. Next, we demonstrate how Ripple can assist in diagnosing such issues by identifying active services and, combined with system-level metrics~\cite{rezvani2024characterizing, jha2022holistic, seo2022nosql, pham2024baro, li2022causal,lin2018microscope, landau2025ebpf,landau2026beyond}, revealing cause of performance degradation.

\begin{figure}[t]
    \centering
    %\upp\upp
    \subfloat[\label{fig:ob-throughput}]{
      \includegraphics[width=0.23\textwidth]{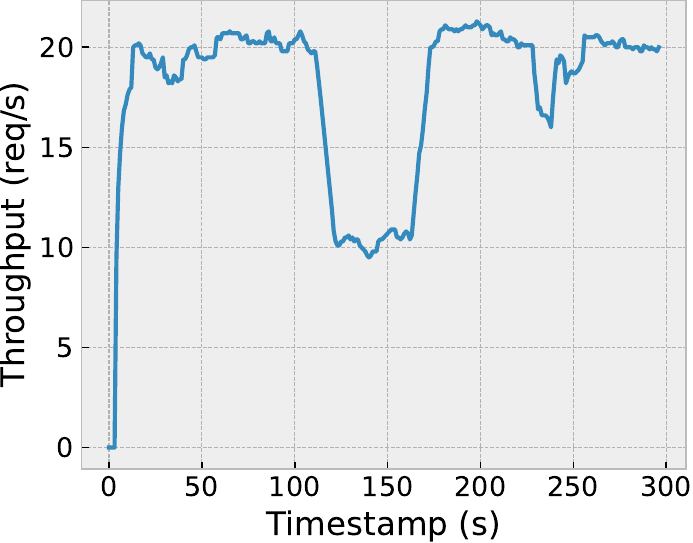}
    }
    \subfloat[\label{fig:productcatalog-rq-time}]{
      \includegraphics[width=0.23\textwidth]{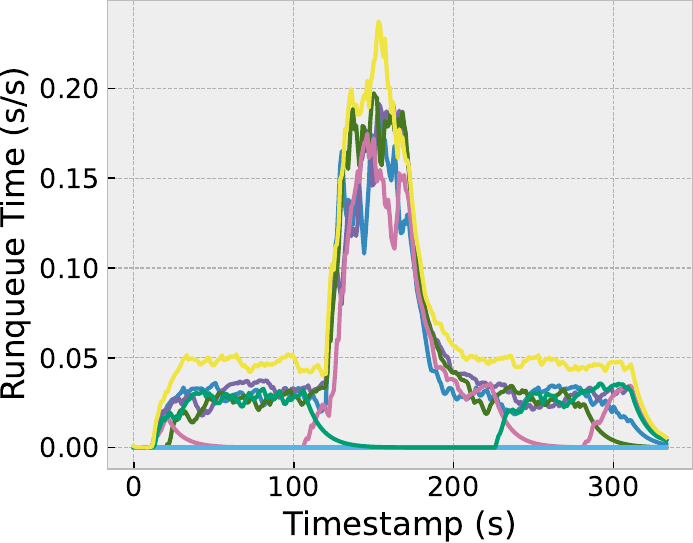}
    }
    \vspace*{-0.2cm}
    %\upp
    \caption{
        (a) \emph{frontend} throughput measured in requests per second; (b) \emph{productcatalog} per-thread runqueue time;
    }
    %\upp
\end{figure}

To illustrate this, we use the Online Boutique microservice benchmark, bootstrapping Ripple from the \texttt{frontend} service. After deploying the microservice benchmark, we initiate a steady workload using its built-in load generator, which continuously invokes the \texttt{frontend} service's \texttt{/product} endpoint. Following approximately 2 minutes of stable execution, we introduce a fault by stressing the \texttt{productcatalog} service using a CPU-intensive \textit{stress-ng} \cite{stress-ng} workload which saturates a full CPU core. This artificial contention degrades the performance of the service, leading to a noticeable decrease in end-to-end throughput observed at the client side. Figure \ref{fig:ob-throughput} illustrates the shift in measured throughput before and after the fault injection.

\begin{figure}[t]
\centering
\includegraphics[width=\columnwidth]{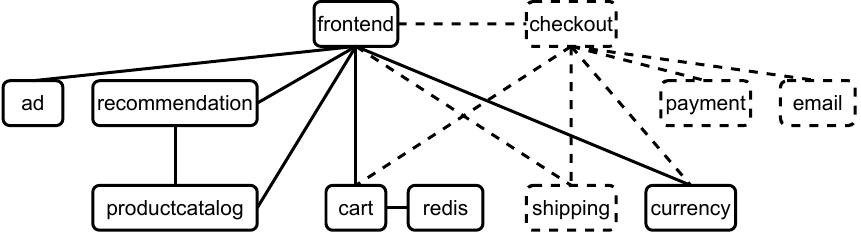}
%\upp
\caption{Service dependency graph of Online Boutique discovered by Ripple. Solid lines indicate active service communication during degradation; dotted lines represent inactive services.}
\label{fig:ob-dependency-graph} \upp
%\upp
\end{figure}

Using Ripple, we extract the dependency graph shown in Figure~\ref{fig:ob-dependency-graph}, which captures the actual run-time communication between services throughout the degraded period. This filtering reduces the scope of analysis, eliminating potential candidates for Root Cause analysis, and enabling operators to concentrate only on the subset of the distributed system that is most likely involved in the observed performance degradation. Leveraging this topology, we examine the services directly connected to the \texttt{frontend} service, revealing a marked increase in CPU runqueue time for the \texttt{productcatalog} threads. This indicates that these threads are having to wait more than usual to be scheduled on a CPU core (due to \textit{stress-ng}), leading to the measured throughput decrease.

In summary, discovering the run-time service topology and narrowing the analysis to inter-dependent services, Ripple enables targeted inspection of affected services. Combined with system-level instrumentation, this integrated approach not only identifies the impacted service but also reveals the underlying cause, in this case, CPU contention, facilitating performance degradation diagnosis in a distributed setting.

\section{Discussion}
\textbf{Protocol Scope:} Due to its design decisions, Ripple currently supports all Layer 5+ protocols built on top of TCP. UDP, on the other hand, presents a distinct set of challenges that require a different approach. In particular, the stateless nature of UDP invalidates the assumption that a discovery identifier only needs to be sent once per connection to minimise overhead. The same UDP socket can be re-used to communicate with multiple destination processes, meaning that there is no persistent connection to which a single discovery identifier can be associated. Unlike TCP, where the sending and receiving kernels maintain state for each connection, UDP maintains no equivalent connection state. Consequently, the same UDP 5-tuple can be handled by different servers over time (e.g. Anycast UDP).

A separate limitation concerns application-layer protocols that establish communication relationships through application-specific logic. Ripple is designed to construct a Layer 4 distributed service dependency graph and therefore does not directly infer dependencies that are established exclusively at higher layers. Using Kafka as an example, when a producer publishes data that is subsequently consumed by a subscriber, this represents a logical communication relationship between the producer and consumer. However, Ripple observes the underlying network communication and consequently includes the intermediate Kafka brokers that forward the data. Thus, the resulting graph represents the network-level communication path, which is producer~$\rightarrow$broker~$\rightarrow$~consumer rather than the logical application-level relationship producer~$\rightarrow$~consumer.

\textbf{State Management:} Running Ripple as a long-lived process while minimising its impact on the observed system requires several conservative approaches to state management and failure handling. For example, Ripple uses an LRU \emph{eBPF} hash map to bound the amount of state maintained by each agent and prevent state from accumulating indefinitely. As a result, state associated with a connection may be evicted from the map over time. If the connection is subsequently observed again, Ripple may therefore treat it as a new connection and transmit another discovery identifier. Similarly, to detect when an identifier has not successfully reached its intended recipient, Ripple agents use a bidirectional communication mechanism. When an agent transmits an identifier for a particular connection, it also expects to receive an identifier over that same connection. If no identifier is subsequently received, the agent can retry the discovery at a later stage.

\textbf{Deployment Considerations:} A critical assumption underlying Ripple's discovery mechanism is that the discovery identifier is preserved as it traverses the network until it reaches the intended receiver. Consequently, if a hardened middlebox strips the experimental TCP Options from a packet, an occurence measured to occur in only $29$ out of $103,000$ paths by Hilal et al.~\cite{yazdani2025poster,hilal2023yarrpbox}, the discovery information will not propagate to the intended receiver. Nevertheless, if such a middlebox administrator wishes to enable Ripple's discovery mechanism within a hardened network, they can configure the relevant middleboxes to allow the single experimental TCP option used by Ripple agents while continuing to block other TCP options.

% \textbf{Program State Management:} The decision to have Ripple run as a long-running process while still minimising its impact on the observed system required the implementation of multiple conservative programming practices. An example of this involves using the LRU \emph{eBPF} hash map to avoid indefinitely accumulating state. Consequently, a connection that was once included in the map might no longer be included, which may result on more than one identifier being sent through that connection. Similarly, to identify when an identifier hasn't been successfully sent to the intended recipient, each Ripple agent employs a two-way communication mechanism wherein an agent sending an identifier for a particular connection also expects to receive an identifier over that same connection. If an identifier is not received over the connection, it can retry at a later stage.

\section{Conclusion}
In this paper, we presented Ripple, a system for constructing accurate, process-level service dependency graphs. Ripple is a run-time approach designed to retrofit onto an already running distributed system that does not require re-compiling or re-deploying an application. Different from existing run-time approaches that passively observe the network metadata exchanged between processes, a method that is susceptible to fail under NAT, Ripple implements a non-disruptive method of injecting metadata onto a TCP packet's header that maintains protocol correctness across host boundaries.

Central to its approach, Ripple incorporates five key design goals: (i) protocol agnosticism, supporting encrypted and non-encrypted TCP-based protocols (e.g. \emph{HTTPS} and \emph{gRPC}); (ii) fine-grained visibility, discovering dependencies at the host$\rightarrow$process level; (iii) userspace dependency mapping, capturing intermediate processes such as load balancers; (iv) graceful deployment, ensuring no disruption even when deployed on a subset of hosts; and (v) non-intrusiveness, operating on unmodified binaries.

To evaluate Ripple, we conducted extensive experiments, comparing it to three state-of-the-art systems (i.e., Caretta, Coroot, and DeepFlow) across three microservice benchmarks and three network configurations (NAT-free, internal-NAT, external-NAT). While other systems failed completely or partially under NAT, Ripple demonstrated its unique robustness, achieving 100\% precision and recall in the majority of external-NAT scenarios and was the only approach to perform consistently across all nine tested environments. Furthermore, it introduced minimal run-time overhead, comparable to the least intrusive methods, and its discovery time scaled primarily with communication patterns rather than the number of services.

To demonstrate Ripple’s practical value, we presented a use case showing how its accurate dependency graphs can narrow the search space for root cause analysis. Despite its strengths, Ripple is currently limited to stateful protocols built on TCP, and extending support to stateless protocols such as UDP is part of future work. Nonetheless, Ripple delivers a reliable, low-overhead, and generalizable approach to service dependency discovery over TCP, enabling critical observability in modern distributed systems regardless of network complexity.

\section*{Acknowledgment}
This work was funded by CETPartnership, the Clean Energy Transition Partnership under the 2024 joint call for research proposals, co-funded by the European Commission
(GA N°101069750) and with the funding organizations FFG
Austrian Research Promotion Agency (Austria), NWO Dutch
Research Council (the Netherlands, Grant EP.1602.24.003), Swedish Energy Agency
(Sweden) and GSRI (Greece).

% \appendix \label{appendix}
% \renewcommand{\thesubsection}{\Alph{subsection}}
% \section*{Appendix: Artifact Availability and Reproducibility}
% The complete source code, functional pipeline, and reproducibility artifacts for Ripple are publicly available in an \emph{online} \verb|GitHub| repository~\cite{repo,repro}. This repository includes the full implementation, software dependencies, and step-by-step instructions required to execute the Ripple’s workflow and reproduce the experimental results presented in this paper.

\appendix \label{appendix}
\renewcommand{\thesubsection}{\Alph{subsection}}
\section*{Artifact Identification and Description}
% This two-page appendix includes the \emph{Artifact Description (AD) and Artifact Evaluation (AE)}, while preserving \emph{double‑blind review guidelines}. Ripple’s complete source code and reproducibility artifact is also released opensource in an \emph{anonymized online} \verb|GitHub| repository~\cite{repo} without violating any anonymization principles.
%\upp\upp
%\subsection*{Artifact Identification and Description}
% \subsubsection*{Artifact Identification and Description}\hfill\\
%\paragraph{Overview of Contributions and Artifacts} 
{\bf \emph{1. Paper's Main Contributions}:} $C_1$: Automated distributed service dependency discovery; $C_2$: Incremental adoption without breaking pre-established connections; $C_3$: Process-level dependency mapping; and $C_4$: Protocol agnostic over TCP.
% \circled{$C_1$}: Automated distributed service dependency discovery; \circled{$C_2$}: Incremental adoption without breaking pre-established connections; \circled{$C_3$}: Process-level dependency mapping; \circled{$C_4$}: Protocol agnostic over TCP.
% \begin{itemize}[label={}]
%     \item $C_1$: Automated distributed service dependency discovery;
%     \item $C_2$: Incremental adoption without breaking pre-established connections;
%     \item $C_3$: Process-level dependency mapping;
%     \item $C_4$: Protocol agnostic over TCP.
% \end{itemize}

\noindent
{\bf \emph{2. Computational Artifact $A_1$}:} The complete source code, functional pipeline, and reproducibility artifacts for Ripple are publicly available in an \emph{online} \verb|GitHub| repository~\cite{repo,repro}. This repository includes the full implementation, software dependencies, and step-by-step instructions required to execute the Ripple’s workflow and reproduce the experimental results presented in this paper. 
% \url{https://github.com/yyuhj46l4y0s7mn/ripple/releases/download/artifact/artifact.zip} 
% provides the instructions and data necessary to reproduce the analysis in Section~\ref{sec:evaluation} pertaining to $C_1$-$C_4$.
%\paragraph{Computational Artifacts}
% \begin{itemize}[label={}]
%     \item $A_1$: \url{https://github.com/yyuhj46l4y0s7mn/ripple/releases/download/artifact/artifact.zip}
% \end{itemize}
\begin{center}
\resizebox{0.8\columnwidth}{!}
{
\begin{tabular}{rll}
\toprule
Artifact ID  &  Contributions Supported &  Related Paper Elements\\
             % &  Supported     &  Paper Elements \\
\midrule
$A_1$   &  $C_1$, $C_2$, $C_3$, $C_4$ & Tables 1-2, Figures 5-9 \\
        % &                                    & Figures 5-9 \\
\bottomrule
\end{tabular}
}
\end{center}

% %%%%%%%%%%%%%%%%%%%%%%%%%%%%%%%%%%%%%%%%%%%%%%%%%%%%%%%%
% \subsubsection*{Artifact Identification}\hfill\\
% %%%%%%%%%%%%%%%%%%%%%%%%%%%%%%%%%%%%%%%%%%%%%%%%%%%%%%%%
% {\bf \emph{1. Computational Artifact $A_1$}:}
%\newartifact
%\paragraph{Relation to Contributions}
%\artrel

% \emph{(a) Relation to Contributions.}
% This artifact $A_1$ provides the instructions and data necessary to reproduce the analysis in Section~\ref{sec:evaluation} pertaining to $C_1$-$C_4$.

\emph{(a) Expected Results.}
%\artexp
Artifact provides instructions for reproducing the service dependency graph results in Section~\ref{sec:results-reconstruction}. End-users of tool can obtain service dependency graphs and scores matching those reported in Section~\ref{sec:evaluation} and Table~\ref{table:reconstruction-results}. Instructions for scalability, overhead, and motivating‑example results (Sections~\ref{sec:results-scalability},\ref{sec:results-overhead},\ref{sec:results-motivating-example}) are also included. While absolute metrics may vary across platforms, the relative performance differences between Ripple and SOTA tools should remain consistent with those presented in the paper.

% The artifact includes the instructions to reproduce the results in Section~\ref{sec:results-reconstruction}. It is expected that the service dependency graph results matches the results obtained in Section~\ref{sec:evaluation}, i.e., Ripple maps the distributed systems with a score identical to the one obtained in Table~\ref{table:reconstruction-results}. 

% The artifact also contains instructions to reproduce the results in Sections~\ref{sec:results-scalability}, \ref{sec:results-overhead} and \ref{sec:results-motivating-example}. These results' variability may be influenced by the platform where Ripple and the tools it is compared with are deployed, however, what should remain, is that when comparing between the same tools in identical conditions, the relative differences should be similar to those observed in these Sections.

%\arttime

\emph{(b) Expected Reproduction Time (in Minutes).}
Reproducing the $36$ experiments in Section~\ref{sec:results-reconstruction} requires $\approx$ \SI{108}{\minute}: \SI{1}{\minute} per experiment run plus \SI{2}{\minute} for environment setup and data collection. The data collection also pertains to gathering datasets provided by SOTA tools 
%(Coroot, Caretta, and Deepflow) 
for comparison. Analysis time is negligible, given we provide scripts to analyse data for each tool.

% For the experiments in Section~\ref{sec:results-reconstruction}, there are 36 experiments in total. Each experiment should run for at least 1 minute, and approximating the time it takes to setup the environments and gathering the datasets provided by the tools we are comparing with (Coroot, Caretta, and Deepflow) to 2 minutes, it should take the user approximately $3 \cdot 36 = 108\text{ minutes}$ to reproduce those results. Given we provide scripts to analyse the data provided by each tool, the time for analysis is negligible.

% For the experiments in Section~\ref{sec:results-overhead}, each approach should run for approximately 30 seconds, and should run 30 times. Therefore it should take a user $30 \cdot 30 \cdot 4 = 3600s = 60 \text{ minutes}$ to reproduce the results. Assuming a kubernetes cluster is already setup, the setup time is negligible given the kubernetes manifests provided by each microservice benchmark. Similarly, the analysis scripts we provide should make the analysis time negligible.

%\artin
\emph{(c) Artifact Setup (incl. Inputs):} \verb|(i) Hardware:|
%\artinpart{Hardware}
The artifact has no strict hardware requirements, although adequate computational resources are recommended for replication purposes. Our setup used $3$ x86\_64 VMs ($2vCPUs$, \SI{8}{\giga\byte} RAM each) with ``private'' IPs for internal communication and ``public'' IPs behind a NAT gateway for external access and forwarding packets to the intended VM.
\verb|(ii) Software:|
The artifact was evaluated on Linux kernel \texttt{6.11.0}  (with BPF/BTF enabled). To reproduce the Kubernetes experiments, each machine should run Kubernetes \texttt{v1.33.0} with \texttt{containerd} runtime and flannel as the the container network interface. For Docker-based experiments, we used \texttt{Docker 28.1.1}. All software dependencies to run Ripple are managed via the provided \texttt{flake.nix} and \texttt{flake.lock} files in the repository~\cite{repo}
% \footnote{\url{https://github.com/yyuhj46l4y0s7mn/ripple/tree/master}}. 
The only prerequisite is the Nix package manager,\footnote{Nix, \url{https://nixos.org/download/}}. A separate flake in the artifact’s root directory handles software dependencies for the experiments. \verb|(iii) Datasets / Inputs:|
Ripple was evaluated using: (a) DeathStarBench\footnote{DeathStarBench, \url{https://github.com/delimitrou/DeathStarBench/}}: social media and media microservice benchmark applications; (b) Online Boutique\footnote{Online Boutique, \url{https://github.com/GoogleCloudPlatform/microservices-demo/tree/main}} microservice benchmark application.
\verb|(iv) Installation and Deployment:|
For reproducing the Kubernetes experiments, all three machines must be part of the same Kubernetes \texttt{v1.33.0} cluster using \texttt{containerd} runtime and \texttt{flannel}. Docker experiments require only \texttt{Docker 28.1.1}. All remaining dependencies are provided by artifact's \texttt{flake} configuration.

\emph{(e) Artifact Execution:}
Main experimental stages are: \textbf{(1) Setup the environment}: For Kubernetes (NAT‑free), set up a cluster with 1 control and 2 worker nodes. For Docker (internal/external NAT), only Docker installed on three machines is required; \textbf{(2) Deploy microservice benchmark}: Use the provided Kubernetes or Docker manifests for deployment; \textbf{(3) Deploy the monitoring tool}: Follow the artifact's \texttt{README} to deploy Ripple, and compared SOTA tools;
%Coroot, Caretta, and Deepflow; 
\textbf{(4) Gather the data}: Ripple outputs a DuckDB database file; the other tools provide data via their UIs for download; and \textbf{(5) Analyse the data}: Place the data in the designated directory and run the provided analysis script to compute scores for each experiment.

\emph{(f) Artifact Analysis (incl. Outputs).}
Each benchmark–environment–tool dataset used for analysis in Table~\ref{table:reconstruction-results} and Figures~\ref{fig:ripple-k8s-reconstruction}–\ref{fig:docker-proxy} should be stored in its corresponding \texttt{data/validation/<benchmark>/<env>/<tool>} directory (\texttt{data/validation/social/k8s/coroot}). 

Running the provided postprocessing script (\texttt{postprocess-coroot.py}) in \texttt{data/validation/<bench>/<env>/} generates a \texttt{post.csv} file listing all nodes and edges in service dependency graph. After processing all tools for a given benchmark–environment pair, the \texttt{compare.py} script outputs the f1‑score, precision, and recall for each tool.

\section*{Artifact Evaluation}
For brevity, this artifact evaluation focuses on Ripple using the online‑boutique benchmark in a NAT‑free Kubernetes cluster ($3$ nodes: $1$ control node \texttt{vm1}, $2$ workers \texttt{vm2}, \texttt{vm3}). Instructions for all other tool–benchmark–environment combinations are in the artifact’s \texttt{README}. The working directory on each VM is \verb|/home/username/ripple|.

% For brevity, this Section focuses on evaluating Ripple on the online boutique microservice benchmark in the NAT-free environment (3 node kubernetes cluster). Nevertheless, the procedure for the remaining tool-benchmark-environment combinations is made available in the artifact's \texttt{README}. We assume a 3 node kubernetes cluster is available, with 1 control node (vm1), and 2 worker nodes (vm2, vm3). Furthermore, we assume the working directory when running the commands on each virtual machine is \verb|/home/username/ripple|.

% \subsubsection*{Computational Artifact $A_1$}\hfill\\
\noindent
{\bf \emph{1. Artifact Setup (incl. Inputs)}:}
%\arteval{1}
%\artin
There are 3 VMs that serve as k8s nodes, and another VM that serves as management node used to interact with k8 VMs with \texttt{kubectl}, \texttt{ssh} and \texttt{scp} commands. This management node is also used to analyse the data produced in the experiments.
Start by unpacking the \href{https://github.com/yyuhj46l4y0s7mn/ripple/releases/download/artifact/artifact.zip}{artifact} in the management node's filesystem, and navigate to the unpacked artifact's root directory:
\begin{lstlisting}[style=terminal]
$ unzip ./artifact.zip -d artifact && \
    cd artifact
\end{lstlisting}

The following command enters a development environment with the software dependencies installed to run the experiments:
\begin{lstlisting}[style=terminal]
$ nix develop --extra-experimental-features nix-command --extra-experimental-features flakes .
\end{lstlisting}

To setup the online boutique microservice benchmark, run:
\begin{lstlisting}[style=terminal]
$ kubectl apply -f "https://raw.githubusercontent.com/GoogleCloudPlatform/microservices-demo/refs/heads/main/release/kubernetes-manifests.yaml"
\end{lstlisting}

To bootstrap Ripple, we need to get the process identifier for one of the online boutique services. For illustration purposes, we will use the adservice pod as our bootstrapping process. List the running online boutique services with: 
\begin{lstlisting}[style=terminal]
$ kubectl get pods -o wide
# NAME                       ... NODE ...
# adservice-dbd9db68f-w4gmb  ... vm1  ...
...
\end{lstlisting}

The output shows that the adservice pod is running on vm1. To get the PID for the services running within the adservice pod:
\begin{lstlisting}[style=terminal]
$ ssh vm1 -t 'sudo crictl inspect "$(sudo crictl ps 2>/dev/null | grep adservice | awk "{print \$1}")" 2>/dev/null | jq ".info.pid"'
# 12345
\end{lstlisting}

The previous command outputs the pid for the adservice entrypoint process, which in this case is \texttt{12345}. We can use this as our bootstrapping PID when starting the Ripple agent on vm1.

% \artexpl{
% Provide instructions for installing and compiling libraries and code. 
% Offer guidelines on deploying the code to resources.
% }

%\artcomp
\noindent
{\bf \emph{2. Artifact Execution}:}
We initiate one Ripple agent per machine, i.e., a Ripple agent is started on vm1, vm2, vm3. Start a ripple agent on the non-bootstrapped machines. On vm2 and vm3 run:
\begin{lstlisting}[style=terminal]
$ sudo RUST_LOG=info nix run github:yyuhj46l4y0s7mn/ripple#ripple -- --machine-id 2
\end{lstlisting}
\upp\upp
%and on vm3, run:
\begin{lstlisting}[style=terminal]
$ sudo RUST_LOG=info nix run github:yyuhj46l4y0s7mn/ripple#ripple -- --machine-id 3
\end{lstlisting}

Next, start the bootstrapped agent with the adservice PID on vm1 (note the additional \verb|--pids| parameter passed into the command): 
\begin{lstlisting}[style=terminal]
$ sudo RUST_LOG=info nix run github:yyuhj46l4y0s7mn/ripple#ripple -- --machine-id 1 --pids 12345
\end{lstlisting}

Upon letting the agents run for approximately one minute, we can terminate each agent with the \texttt{SIGINT} signal (\texttt{Ctrl+C}). A duckdb database file of the form \texttt{<timestamp>.db3} should now be available in the relative \texttt{./data} directory.

%\artout
{\bf \emph{3. Artifact Analysis (incl. Outputs)}:}
After execution from previous steps, each VM (vm1, vm2, vm3) will contain a ducdkb database file in \texttt{/home/username/ripple/data}. Copy these files from the k8s nodes to the management node. For example, from the management node's artifact root directory, the following command copies vm1's data to the expected location:
% Resulting from the previous execution, vm1, vm2, and vm3 should each have a ducdkb database file available at \texttt{/home/username/ripple/data}. Copy each file produced in the k8s node to the management node. For example, assuming the current working directory in the management node is the unpacked artifact's root directory, the following command would copy the data generated by ripple in vm1 to the expected location for this experiment:
\begin{lstlisting}[style=terminal]
scp "vm1:/home/username/ripple/data/2026-02-03T17:25:55.425906153+00:00.db3" "./data/validation/online-boutique/k8s/ripple/vm1.db3"
\end{lstlisting}

Having run the previous command for the 3 VMs, the \verb|data/validation/online-boutique/k8s/ripple| directory should now include the database files of each ripple agent, i.e., \verb|vm1.db3|,  \verb|vm2.db3| and \verb|vm3.db3|. The data analysis is performed by the \verb|postprocess-ripple.py| script which creates the service dependency graph from the database files, and outputs the graph data into a \verb|post.csv| file. As such, running:
\begin{lstlisting}[style=terminal]
python data/validation/online-boutique/k8s/postprocess-ripple.py data/validation/online-boutique/k8s/ripple
\end{lstlisting}
creates the \verb|post.csv| file in the \verb|data/validation/online-boutique/k8s/ripple| directory, which contains data in the following format:
\begin{lstlisting}[style=terminal]
$ cat data/validation/online-boutique/k8s/ripple/post.csv
# src,dst
# adservice,frontend
# cartservice,checkoutservice
# ...
\end{lstlisting}

The procedures to reproduce the data for the remaining combinations of benchmark-environment-tool, are also available in the \verb|README.md| file, in the unpacked artifact's root directory.

% \artexpl{
% \begin{itemize}
%     \item Provide a description of the expected results and a methodology for evaluating these results. 
%     \item Explain how the expected results from the experiment workflow correlate with the contributions stated in the article. 
%     \item For example, if the article presents results in a figure, the artifact evaluation should also produce a similar figure, depicting the same generalizable outcome. Authors must focus on these aspects to reduce the time required for others to understand and verify an artifact.
% \end{itemize}
% }

% \bibliographystyle{ACM-Reference-Format}
\bibliographystyle{IEEEtran}
\bibliography{bibliography.bib}

\end{document}